# On-Chip Vectorial Structured Light Manipulation via Inverse Design

Xiaobin Lin[1†], Maoliang Wei[1†], Kunhao Lei[1], Zijia Wang[1], Chi Wang[1], Hui Ma[1], Yuting Ye[2], Qiwei Zhan[1], Da Li[1], Shixun Dai[3], Baile Zhang[4], Xiaoyong Hu[5], Lan Li[2], Erping Li[1*], Hongtao Lin[1,6*]

[1]State Key Laboratory of Brain-Machine Intelligence, Key Laboratory of Micro-Nano Electronics and Smart System of Zhejiang Province, College of Information Science and Electronic Engineering, Zhejiang University, Hangzhou 310027, China.

[2]Key Laboratory of 3D Micro/Nano Fabrication and Characterization of Zhejiang Province, School of Engineering, Westlake University, Hangzhou 310024, China

[3]Laboratory of Infrared Materials and Devices, Ningbo University, Ningbo 315211, China

[4]Division of Physics and Applied Physics, School of Physical and Mathematical Sciences, Nanyang Technological University, Singapore 637371, Singapore.

[5]State Key Laboratory for Mesoscopic Physics, Frontiers Science Center for Nano-optoelectronics, School of Physics, Peking University, Beijing 100871, China.

[6]MOE Frontier Science Center for Brain Science & Brain-Machine Integration, Zhejiang University, Hangzhou 310027, China.

*Corresponding author. Email: hometown@zju.edu.cn, liep@zju.edu.cn

†These authors contributed equally to this work.

## Abstract

On-chip structured light, with potentially infinite complexity[1-3,1-3], has emerged as a linchpin in the realm of integrated photonics. However, the realization of arbitrarily tailoring a multitude of light field dimensions in complex media remains a challenge[1,4-6]. Through associating physical light fields and mathematical function spaces by introducing a mapping operator, we proposed a data-driven inverse design method to precisely manipulate between any two structured light fields in the on-chip high-dimensional Hilbert space. To illustrate, light field conversion in on-chip topological photonics was achieved. High-performance topological coupling devices with minimal insertion loss and customizable topological routing devices were designed and realized. Our method provides a new paradigm to enable precise manipulation over the on-chip vectorial structured light and paves the way for the realization of complex photonic functions.

# Main Text

Beyond the traditional transverse electric-magnetic state, on-chip structured light can adopt radial or azimuthal polarization and possess angular momentum into integrated photonics[3, 6-8]. Through arbitrarily tailoring of the degrees of freedom (DoFs) of optical fields on chip[2, 9], it is paving the way for advancements in optical communication[10-11], quantum optics[12-13], and laser technology[7, 14-15]. However, due to the complexity of the vectorial light field, the absence of a unified paradigm renders the accomplishment of arbitrary on-chip vectorial structured light conversion a substantial challenge[7, 14-15].

One noteworthy aspect of on-chip structured light is the topological interface modes in the valley photonic crystal (VPC), garnering significant interest for their attributes, such as reflectionless transmission[16-17], topological protection[18-20], and valley polarization[21-22,21-22]. Furthermore, topological photonic insulators accommodate a diverse type of structured light, extending from topological states to topologically trivial states, and from single-mode topological interface states to multimode topological interface states[23-24,23-24]. This diversity means more complex and diverse light field. The precise manipulation of arbitrary vectorial light fields in topological photonic insulators, holds promise for its progression in various optical applications[25]. Yet, the complexity and diversity inherent to structured light of the topological photonics pose a significant challenge for traditional design methods[26], which typically offer a limited search space[27-28].

Inverse design, capable of augmenting the design space, enables unprecedented control of light, opening new avenues for this challenge. It has demonstrated its design flexibility in various fields including wavelength demultiplexers[29-30], computational structures[31-32], and particle accelerators[33], particularly revealing distinct advantages in light field conversion[34-36]. Earlier studies have predominantly homed in on the manipulation of light's amplitude and phase[34], sidelining the vectorial information of the light field. While recent studies have delved into vortex beams in free space, they predominantly hover within the domain of fundamental modes[37]. Given the profound complexity associated with arbitrary on-chip structured light, a unified paradigm remains conspicuously absent. Furthermore, these methods leaned heavily on computational techniques like time-domain finite-difference[38] or finite-element approaches[39], demanding substantial computational resources, particularly for structures of smaller dimensions.

Here, we synergized data-driven and physics-driven approaches to devise a inverse design method. This method established an association between physical light fields and mathematical function spaces by introducing a mapping operator that links any arbitrary structured light field under a quantum-inspire framework. Alongside, we integrated the Discontinuous Galerkin (DG) method and robust-inverse design optimization strategies to enhance and refine our design. Several topological photonic devices exhibiting high performance have been designed and fabricated, including topological coupling devices with minimal insertion loss, and customizable topological routing devices. Our approach not only allows for the manipulation of light polarization but also precisely controls other DoFs of structured light, including phase, wavelength, and spatial distribution. This capability enables the design of unprecedented topological photonic devices, such as broadband phase shifters, wavelength division multiplexers, and free-space couplers. These devices are indispensable for the construction of future quantum networks and are challenging to achieve using conventional methods. Inverse design has provided a significant boost to the domain of integrated photonics, unfolding more extensive application possibilities of the on-chip structured light.

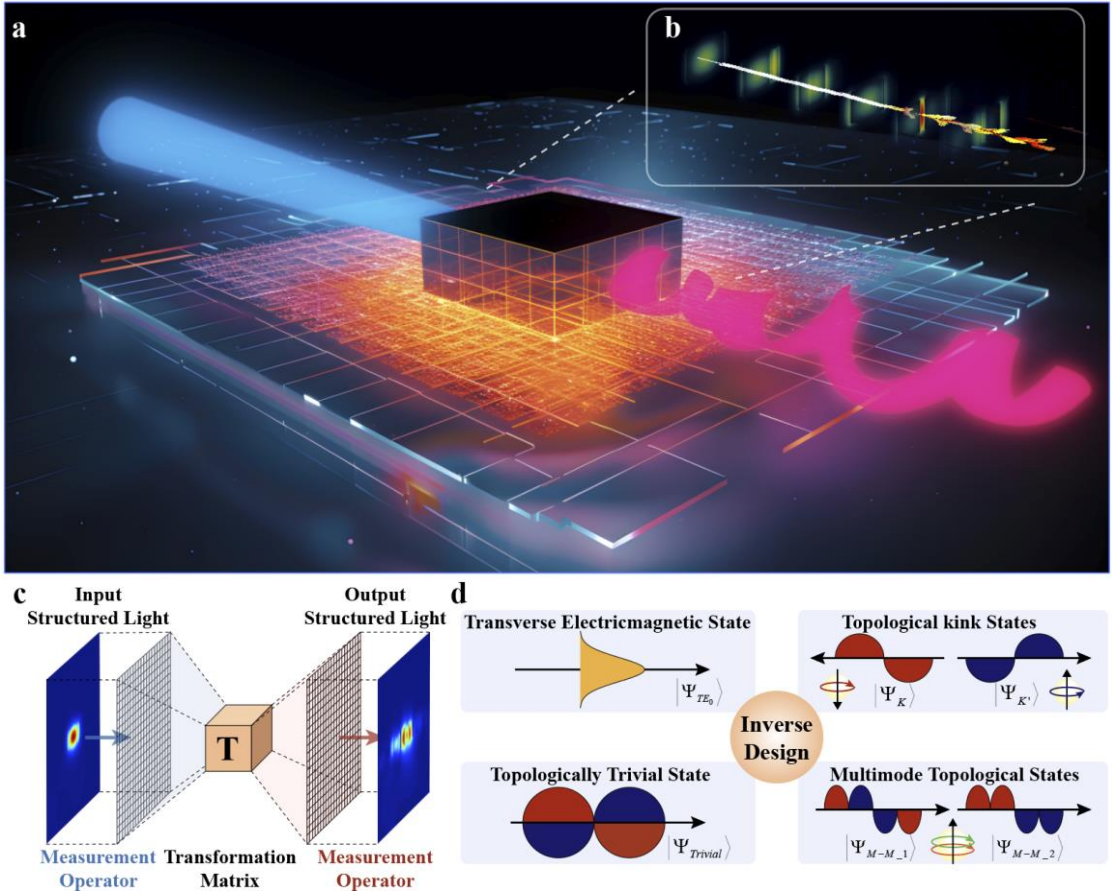

Fig.1. Inverse design to achieve arbitrary on-chip vectorial structure light conversion. (a) Schematic of the arbitrary on-chip structured light conversion. (Inset) Black area indicates the inverse-designed structure. (b) Light field evolution from TE0 mode to pseudo-spin mode. (Inset) Orange arrows represent wave vectors. (c) Flow chart of vector-based inverse design method. (d) Classification of structured light in topological photonics.

**Vector-based inverse design for arbitrary vectorial structured light manipulation**

An arbitrary on-chip vectorial structured light field, irrespective of its complexity, can be succinctly expressed in the quantum state[5]:

$$|\Psi\rangle = \sum_i \sum_j C_{i,j} |e_i\rangle_A |u_j\rangle_B \tag{1}$$

here $|\Psi\rangle$ denotes an unnormalized superposition of different spatial and polarization states, signified by $|e\rangle_A$ and $|u\rangle_B$ respectively. The coefficients $C_{i,j}$ are complex quantities whose squared magnitudes reveal the probability of finding the light field in a corresponding blend of two DoFs.

This conceptualization enables us to perceive each vectorial structured light field as a vector situated in a high-dimensional complex vector space, also known as a Hilbert space. Each dimension of this space corresponds to a plausible state or configuration of the light field. The coordinates of the vector within this space are determined by the complex coefficients that define the superposed state of the light field, encapsulating crucial information about the field. These coefficients denote properties linked to the selected basis, such as the light field's polarization and spatial distribution.

Transformations among different structured light fields can be perceived as operations on these vectors within the Hilbert space, represented as:

$$|\Psi_{out}\rangle = \hat{\mathbf{U}}|\Psi_{in}\rangle \tag{2}$$

here $\hat{\mathbf{U}}$ stands for a unitary operator that encompassing the physical processes during light propagation or interaction with optical elements, essentially encoding the physical principles of the light-medium interaction. The operator's exact form is contingent upon the specific physical process involved.

However, directly manipulating a continuous, infinite-dimensional Hilbert space is a complex task, particularly when the light field may be described by transcendental functions. A logical strategy is to execute dimensionality reduction by judiciously setting the DoFs of the low-dimensional initial state [1]. For this purpose, we define a measurement operator $\mathbf{L}$, representing our physical quantity of interest. As shown in Fig. 1c, for light field measurements, we opt for a plane along the direction of propagation and measure the vector electric field information at different points on the light field. This process reduces a continuous three-dimensional structured light field $|\Psi\rangle$ to an equivalent complex matrix $\mathbf{E}$, which explains the measurement results at various points of the light field:

$$\mathbf{E} = \mathbf{L}|\Psi\rangle \tag{3}$$

here, the operator $\mathbf{L}$ can be seen as a functional. By selecting specific DoFs, we choose a finite subset from this infinite-dimensional Hilbert space and map this subset to a finite-dimensional complex matrix $\mathbf{E}$.

The evolution of the light field is now illustrated as a transformation matrix between two complex matrices, corresponding to the initial and final states of the light field:

$$\mathbf{E}_{out} = \mathbf{T} \cdot \mathbf{E}_{in} \tag{4}$$

here, $\mathbf{E}_{out}$ and $\mathbf{E}_{in}$ are complex matrices corresponding to the input and output structured light fields, and $\mathbf{T}$ is the mapping matrix, which describes how to convert between the two structured light fields in the finite-dimensional Hilbert space. Although this procedure solely involves 2D cross-sectional data, the appropriate choice of a suitable mesh[40] and our adherence to Maxwell's equations assure the complete recovery of the original three-dimensional light field without any information loss.

We leverage the quantum-inspired framework in inverse design to enable the transformation of any vectorial structured light (Fig.1a). The inverse design optimization algorithm is employed to design an inhomogeneous permittivity distribution within a predefined compact region to yield the mapping matrix. Here, the transfer function of the mapping matrix is kept as a constraint, and with each iteration, the relative permittivity asymptotically approaches our goal.

Traditional design methods for light field manipulation typically employ the mathematical characterization of light fields for theoretical description or problem-solving. However, such approaches frequently struggle with complex light field, such as the pseudo-spin modes found in the valley topological photonic crystals[41]. Even more, the on-chip structured light in VPC can be compared to the vortex light in free space (discussed in supplementary text, section S1). These transformations may involve transitions between several different types of structured light in topological photonic, including modes in standard line waveguides, pseudo-spin modes, modes in topologically trivial states, and modes in multimode topological interface states (Fig.1d). Often possessing inherent complexity and properties beyond the scope of classical optics, these complex vectorial light field make it difficult to derive a rigorous and usable analytical expression. In contrast, as shown in Fig. 1b, our method expresses the relationship between different light field states in an infinite-dimensional mathematical function space by constructing a mapping matrix, which allows us to establish a unified paradigm to realize arbitrary structured light operations (Supplementary Information section 2.1) .

To boost computational precision, we employ the discontinuous Galerkin method[40] for adaptive grid modeling, enabling swift calculations. Concurrently, a robust inverse design framework is incorporated (Supplementary Information section 2.2), thereby enhancing the design stability. This framework fortifies the design's tolerance, accommodating potential process errors, and significantly heightening the design's practicability. By inversely designing the mapping matrix, we circumvent intricate physical operations, effectively achieving the conversion of optical field modes. Our novel framework exhibits extensive applicability, handling a multitude of light field dimensions, extending beyond simple transverse plane spaces (2D) to complex structures (3D space), and even beyond classical waves to quantum-structured light.

**Arbitrary vectorial structured light converters for dual-port systems**

Efficient coupling from the lowest order bound mode of the single-mode waveguide to the topological interface mode hinges on the generation of vortex fields[41]. The VPC waveguide explored here rely on a photonic crystal that emulates Haldane model, where holes arranged in a honeycomb structure (a=630 nm), as shown in Fig.2A. Breaking inversion symmetry (R=296 nm, r=133 nm) lifts the degeneracy of the Dirac cone for transverse-electric-like modes at the K/K' point in the Brillouin zone and opens a bandgap between 1910 and 2070 nm. Due to the valley Hall effect, light travels as the valley-polarized topological kink states at the interface of two honeycomb VPCs and carries orbital angular momentum (OAM) chirality with valley dependence (Supplementary Information section 1). Previous works have achieved coupling through line defect waveguides[42-43], multimode waveguides[41], or asymmetric connections[44-45], relying on intuition. However, these methods have faced shortcomings such as high insertion loss (IL), large footprint.

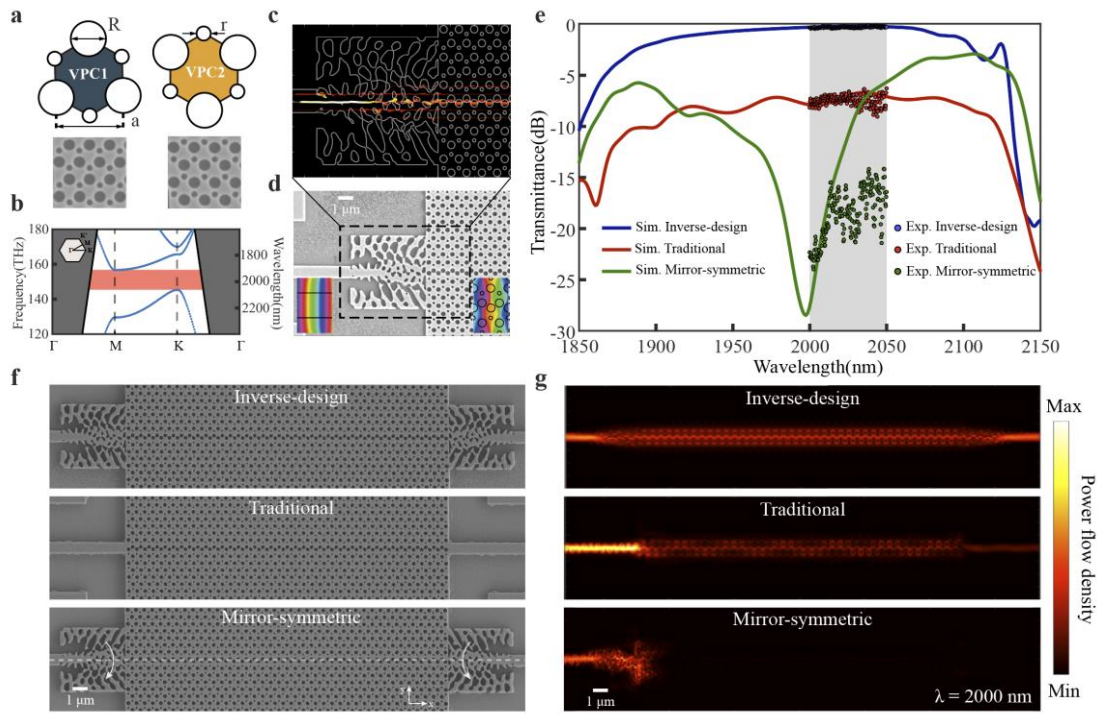

Fig.2. Mode conversion from $TE_0$ mode to pseudo-spin mode (VPC Coupler). (a) Details of two kinds of VPC unit cells (a=630 nm, R=296 nm, r=133 nm). (b) Bulk band both for VPC1 and VPC2. (c) Wave-vector distribution of VPC coupler based on inverse design. (d) Top-view SEM image of the sample. (Inset) $H_z$ phase of the $TE_0$ mode and the pseudo-spin mode. (e) Simulated (Line) and measured (dot) transmission spectra of different connections. Inverse design connection (Blue); Traditional single-mode waveguide connection (Red); Mirror-symmetric connection (Green). (f) Top-view SEM image of different connections. (g) Energy flow distribution of different connections at the wavelength of 2 μm.

By employing our method mentioned above, an inverse design structure (5*5 μm²) that satisfied the mapping matrix relationship between $TE_0$ and topological interface mode was designed. The energy flow and wave vector distribution of the device at the wavelength of 2 μm were simulated, as illustrated in Fig. 2c, showing the efficiently conversation from the $TE_0$ mode to the topological interface mode in the VPC.

The device was fabricated on a chalcogenide photonics platform[46], and the scanning electron microscope (SEM) image of the device was displayed in Fig. 2d. We connected the inverse design structures on both sides of the VPC to measure the transmission of the device. The inverse design device achieved an IL of 0.19 dB (@2033 nm) and a bandwidth of 158 nm (1929~2083 nm, IL<1 dB), as shown in Fig. 2e (blue line, simulation results). The experimental results (Fig. 2e, blue dot) were consistent with the simulations, realizing IL of 0.20 dB (@2023 nm) and bandwidth greater than 50 nm (2000~2050 nm, limited by laser bandwidth). The energy flowed from the input single-mode waveguide into the topological photonic crystal and then into the output single-mode waveguide almost without reflection at 2 μm (Fig. 2g, "Inverse-Design").

We further compared the case of the VPC directly connected with the single-mode waveguide (Fig. 2f). The mismatch in the light field resulted in significant IL (IL=6.68 dB, @2035 nm) and induced oscillations in the transmission spectrum of the device (Fig. 2e, red line, simulation result). Standing waves were observed in both the single-mode waveguide and the photonic crystal based on the simulation results of the energy flow, highlighting marked energy reflection (Fig. 2g, "Traditional"). Furthermore, the bulk-edge correspondence theorem guarantees the presence of two degenerate counterpropagating states localized at the domain wall, demonstrating different chiral symmetry in real space. By mirroring the inverse design structured along the x-axis, we generated a light field with chiral symmetry to the previous mode, yielding a peak extinction ratio (ER) of 28.40 dB at 1997 nm (Fig. 2e, green line, simulation result). This phenomenon well verified the photonic valley–chirality locking property. As a result, we discerned that the energy predominantly resided in the input waveguide, inhibiting its transfer to the photonic crystal (Fig. 2g, "Mirror-Symmetric"). Our experimental findings robustly supported this observation, with an ER of 24.00 dB recorded at 2002 nm (Fig. 2e, green dot). Furthermore, we identified that the mirror-symmetric inverse design structure showcased remarkable coupling efficiency, especially at wavelengths proximate to the band edge. This behavior can be attributed to the modes in the topological photonic crystal transitioning from the topological to the trivial states gradually (Supplementary Information section 3).

Beyond the conventional topological interface modes, other structured light forms hold significant relevance in topological photonics. Our method facilitates intricate light field manipulations, allowing precise structured light transformations for the trivial state in topological photonic crystals. Expanding on the previously mentioned VPC waveguide, we modified the adjacent two rows of holes at the interface, transitioning from small to large holes, thereby crafting a large-area VPC waveguide. These large-area VPC waveguides can be perceived as VPC waveguide with defects. Besides the existing topologically non-trivial interface states, we discovered a novel trivial band at the low frequency. Our method could also achieve efficient conversion of trivial states in photonic crystals while suppressing the excitation of topological states. ((Supplementary Information section 5).

Traditionally, the majority of VPC devices could only operate at a single mode. This limitation largely hindered their extensive application in optical communication, lasers, and quantum optics. To overcome this issue, the multimode VPCs with a large Chern number were introduced[47-48], enabling greater channel capacity. However, existing devices of this type still faced challenges with high IL and mode mixing. Utilizing the Stampfli-triangle photonic crystal, we have constructed a multimode topological waveguide. Remarkably, within the same frequency band, the device can accommodate two distinct topological interface state modes. Our inverse-design method enables the efficient conversation from the $TE_0$ mode to any specified topological interface mode, with the suppression of other modes ((Supplementary Information section 6). This novel approach offers an unrivaled capabilities of light field modulation and versatility, unlocking pioneering design and deployment opportunities for valley topological photonic crystal devices and making significant strides in the on-chip topological photonics.

**Vectorial structured light converters and routers for multi-port systems**

Not limited to arbitrary structured light manipulation in dual-port systems, our algorithm is also applicable to multi-port systems. Besides, the light is essentially transmitted in optical vortex-like modes carrying spin angular momentum, which has huge potential in routing and optical operations. Previous works have achieved the beam routing function through a sub-wavelength micro-disk[41], precisely designed interface[49]. However, these approaches consistently face hurdles, such as heightened IL, a compulsion to tweak the system's geometry, and a lack of flexibility in routing functionalities. Here, we further explore the conversion from $TE_0$ mode to different the valley-chirality locking interface states in a multi-port system to achieve an arbitrary routing function (Fig.3a).

In the VPC, the direction of light propagation depends on the chirality of the valley pseudospin. For the crossing route, the input signal from the up/bottom single-mode waveguide (Port.1/2) passed through the inverse design region and was converted into the kink state bounded to K'/K valley ((Supplementary Information section 3), which was output from the bottom/up port of the photonic crystal (Port.4/3). The mapping relationships was defined as:

$$\begin{vmatrix} \mathbf{E}_{K'} \\ \mathbf{E}_{K} \end{vmatrix} = \mathbf{T} \cdot \begin{vmatrix} 0 & 1 \\ 1 & 0 \end{vmatrix} \cdot \begin{vmatrix} \mathbf{E}_{port\_1} \\ \mathbf{E}_{port\_2} \end{vmatrix} \tag{5}$$

here $\mathbf{E}_{K'}$ and $\mathbf{E}_{K}$ were the complex matrices corresponding to the ideal the kink state. $\mathbf{E}_{port\_1}$ and $\mathbf{E}_{port\_2}$ were the complex matrices corresponding to the single-mode signal input at the port 1 and 2. The energy flow distribution of the crossing route at a wavelength of 2 μm was simulated, demonstrating that light was cross transmitted into different VPC waveguides (Fig.3d). Fig.3f show the simulated transmission spectrum of different output ports of the device when light was input from the Port.1. The IL was 0.22 dB (@2051 nm, S41, red line) and the ER was 53.88 dB (@2046 nm, S31, blue line). Validating these findings with our experimental tests, we observed an average IL of 0.45 dB (measured over 2000~2050 nm, red dot) and an ER of 48.42 dB (@2045

nm, blue dot).

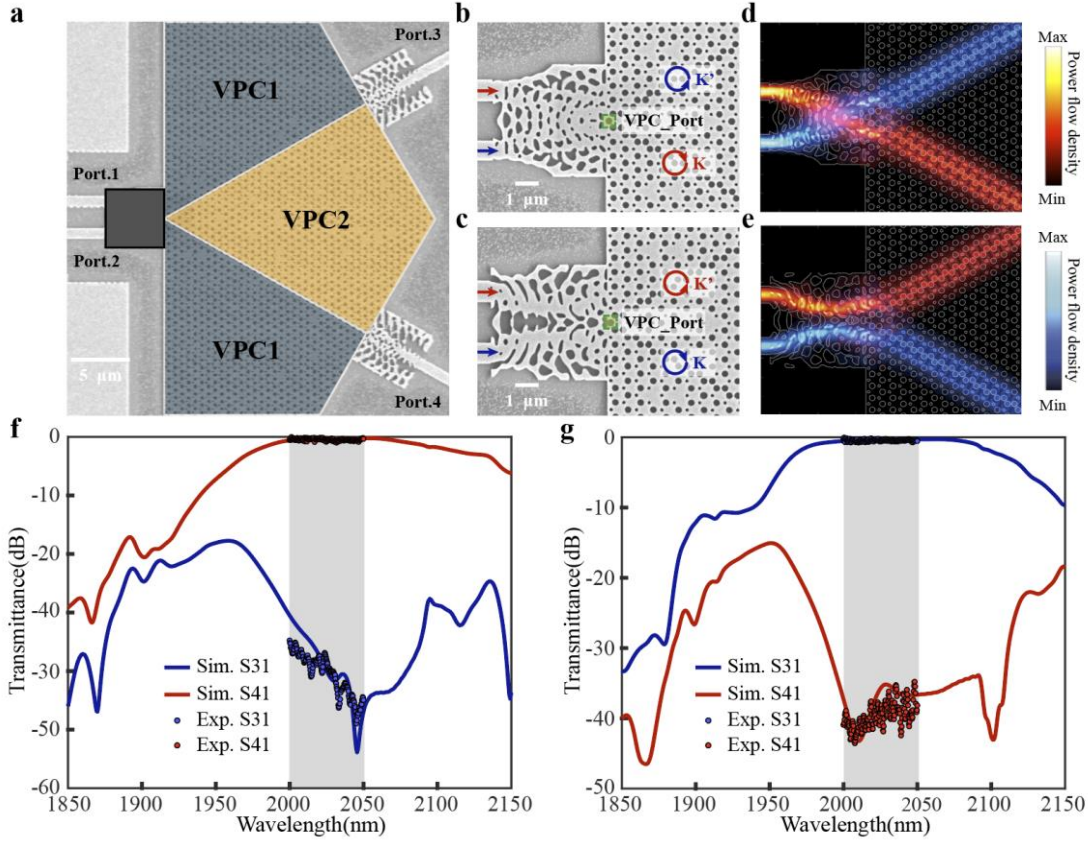

Fig.3. Mode conversion from $TE_0$ mode to arbitrary pseudo-spin mode (VPC router). (a) Top-view SEM image of the VPC router. (b and c) High-resolution top-view SEM image of the crossing route and the parallel route. (d and e) Energy flow distribution of the crossing route and the parallel route at the wavelength of 2 μm. (f and g) Simulated (Line) and measured (dot) transmission spectra of the crossing route and the parallel route.

For the parallel route, the function of the device was the opposite of the cross-routing (Fig.3c), and our objective function was written as:

$$\begin{vmatrix} \mathbf{E}_{K'} \\ \mathbf{E}_{K} \end{vmatrix} = \mathbf{T} \cdot \begin{vmatrix} 1 & 0 \\ 0 & 1 \end{vmatrix} \cdot \begin{vmatrix} \mathbf{E}_{port\_1} \\ \mathbf{E}_{port\_2} \end{vmatrix} \tag{6}$$

the $TE_0$ signal input from the top (Port.1) was converted into the kink state bounded in K' valley and the $TE_0$ signal input from the bottom (Port.2) was converted into the kink state bounded in K valley (Fig.3e). In a manner consistent with our earlier examinations of the cross route, we simulated the transmission spectrum from various ports of the device, with light being sourced from Port.1. Remarkably, the device executed a comprehensive parallel routing function, recording an IL of 0.28 dB (@2045 nm, S31, blue line) and an ER of 42.80 dB (@2010 nm, S41, red line). The empirical findings from our experiments were closely aligned with our simulations, producing an average IL of 0.44 dB (measured over 2000-2050 nm, blue dot) and an ER of 43.17 dB (@2008

nm, red dot). Our research confirms the effective conversion of $TE_0$ light to arbitrary pseudo-spin lights in a multi-port system. The efficacy of our device stands unparalleled in current records, paving the way for innovative strategies in the realm of topological photonic crystal devices.

In addition to the conversion of $TE_0$ to various valley-polarized topological kink states, we delve deeper into the applicability of our method for transitions among valley-polarized topological kink states themselves. The routing within the VPC is also a focus of research due to its applications in optical communications[50], quantum technologies[51], and optical detection[52]. In the previous work, the realization of the routing of the VPC beam splitter has depend on the selection of the interface or the size of the structural shape, limited by valley–chirality locking property and inherent light field mismatch. They all had the disadvantages of high IL and fixed function, limiting the development of topological photonic devices. We apply our inverse design approach to the conversion of arbitrary topological kink states.

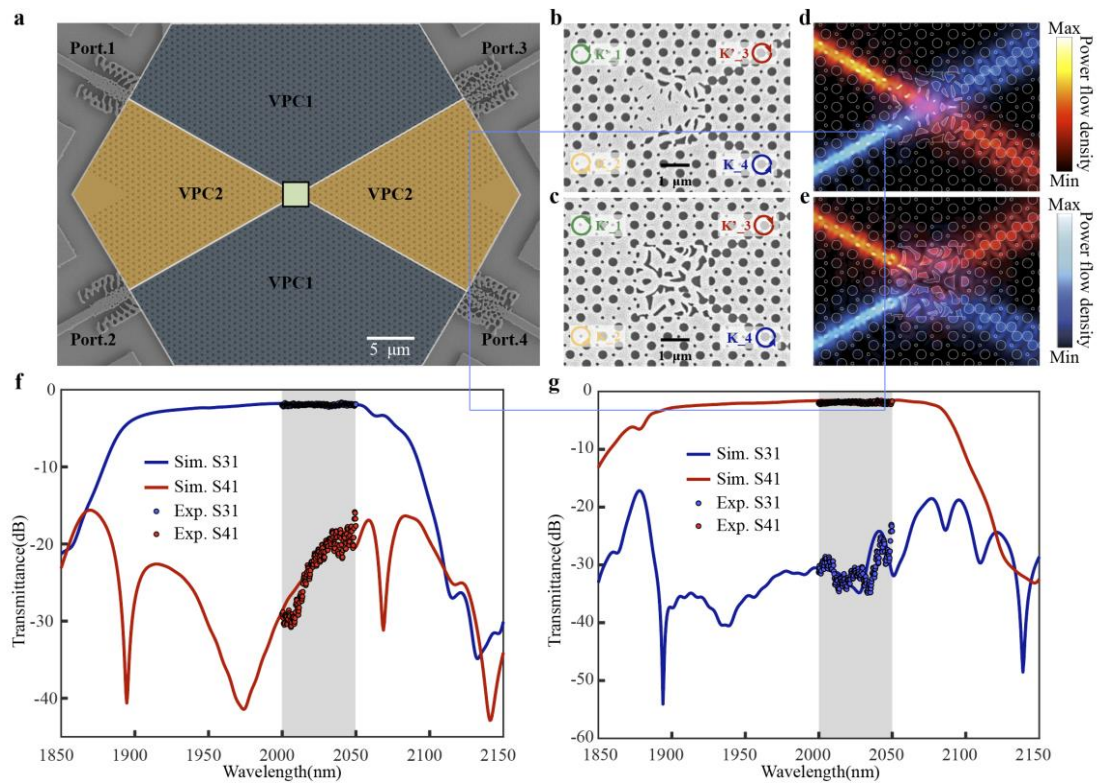

Fig.4. Mode conversion from arbitrary pseudo-spin mode to arbitrary pseudo-spin mode (VPC router). (a) Top-view SEM image of the VPC router. The structural parameters (a=650 nm, R=340 nm, and r=103 nm) was adjust to ensure that different pseudo-spin states have the same bandwidth. (b and c) High-resolution top-view SEM image of the crossing route and the parallel route. (d and e) Energy flow distribution of the crossing route and the parallel route at the wavelength of 2 μm. (f and g) Simulated (Line) and measured (dot) transmission spectra of the crossing route and the parallel route.

Notably, the small feature structure and the existence of periodicity in the VPC lead to finer mesh division and larger simulation area in the computational process. Here, our inverse design method relies on complex matrices corresponding to the ideal vector electric fields of the input and output ports within the boundaries of the design region, without the need to simulate the entire topological photonic crystal device, enabling decentralized inverse design. The mapping function was written as:

$$\begin{vmatrix} \mathbf{E}_{K'\_3} \\ \mathbf{E}_{K\_4} \end{vmatrix} = \begin{cases} \mathbf{T}_{cross} \cdot \begin{vmatrix} 0 & 1 \\ 1 & 0 \end{vmatrix} \cdot \begin{vmatrix} \mathbf{E}_{K'\_1} \\ \mathbf{E}_{K\_2} \end{vmatrix} & cross\ route \\ \mathbf{T}_{parallel} \cdot \begin{vmatrix} 1 & 0 \\ 0 & 1 \end{vmatrix} \cdot \begin{vmatrix} \mathbf{E}_{K'\_1} \\ \mathbf{E}_{K\_2} \end{vmatrix} & parallel\ route \end{cases} \quad (7)$$

here $\mathbf{E}_{K\_i}$ or $\mathbf{E}_{K'\_i}$ were the complex matrices corresponding to the ideal valley-polarized topological kink states bounded in K or K' valley at the $i$th port. To bolster integration capabilities, our design area was minimized to 2.5 * 2.5 μm$^2$, depicted in Fig.4a. Fig. 4b and 4c highlight the SEM structures of two distinct devices tailored for specific functionalities. The energy flow distribution of these devices at the 2 μm wavelength was simulated to analyze their performance metrics. As depicted in Fig. 4d, the kink state bounded to K' valley from the initial interface state (small hole to small hole, K'_1) was transformed into the kink state bounded to K valley, which was composed of large hole to large hole (K_4). In parallel, the kink state bounded to K valley, introduced via the VPC port_2 (K_2), was modified into the corresponding kink state bounded to K' valley (K'_3) and subsequently relayed from the topmost port. Fig.4f presents the simulated transmission spectrum across different device ports, with Port.1 serving as the light source. Our device registered an IL of 1.79 dB (@ 2000 nm) coupled with an ER of 39.40 dB (@1974 nm). Experimental result echoed these findings, recording an average IL of 1.94 dB (2000~2020 nm) and an ER of 28.75 dB (@2006 nm), underscoring the alignment between experimental and simulated outcomes. For the parallel trajectory illustrated in Fig.4e, the initial interface state's (K'_1) was converted into another kink state bounded to K' valley (K'_3). Simultaneously, the kink state (K_2) is converted into its counterpart kink state (K_4). Our simulation data, detailed in Fig. 4f, recorded an IL of 1.49 dB (@2054 nm) and an ER of 38.41 dB (@1939 nm). Experimental results agreed well these findings, yielding an IL of 1.52 dB (@2040 nm) and an ER of 33.08 dB (@2033 nm), underscoring the consistency between simulated and measured results. Though the contraction of the design area has an adverse effect on device performance, the comprehensive routing function remains intact. Furthermore, we conducted digital signal testing on the device to verify its signal transmission capabilities (Supplementary Information section 4). The devices highlighted above effectively underscore our prowess in achieving efficient transitions between intricate on-chip structured light conversion.

**Advanced DOFs Manipulation for Topology Quantum Circuits**

Previous studies have experimentally verified topologically protected two-photon states[53], valley quantum photon circuits[51], and excitation of on-chip polarized quantum entanglement[54], demonstrating the potential of topological photonics in quantum communication. Our design of low loss dual-port and multi-port vectorial structured light converters and routers could ensure the ultra-low-loss transmission of signals, while supporting complex optical functions and the protection of quantum states, greatly advancing the ability to build stable, reliable, and efficient quantum communication systems. However, to meet the demands of advanced topology quantum circuits for complex functions and high performance (Supplementary Information section 7), unprecedented topological photonic devices, such as broadband phase shifters, wavelength division multiplexers, and free-space couplers are needed. These devices are indispensable for the construction of future quantum networks and are challenging to achieve using conventional methods. In addition to polarization, the vector-based inverse design framework should also need to consider the regulation of other dimensions such as wavelength, phase, and so on, to achieve more extensive and complex optical functions.

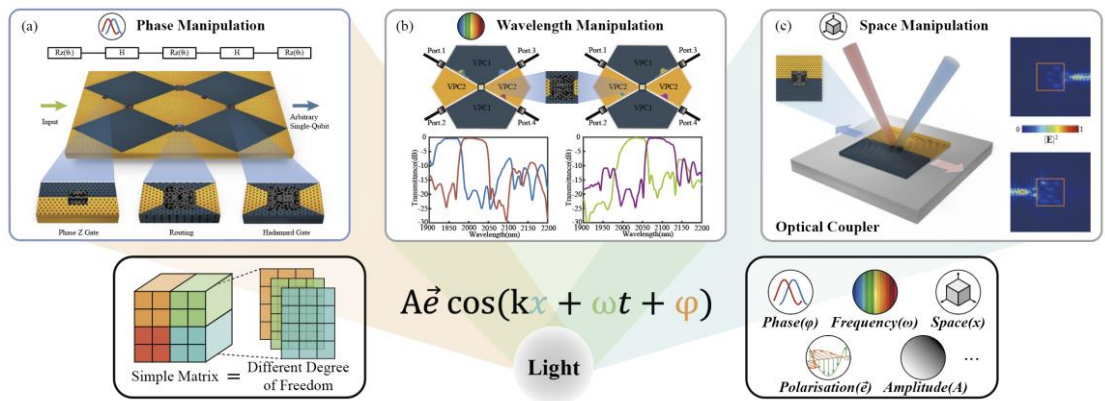

Fig.5. Manipulation of advanced DoFs. (a) Phase manipulation for quantum logic gate design; (b) Wavelength manipulation for WDM device design; (c) Space manipulation for free-space coupler.

Light is commonly described as an electromagnetic wave, characterized primarily by its wavelength, amplitude, phase, and polarization. These parameters form the framework for describing the basic physical behavior of light. In fact, light possesses multiple DoFs, each representing a dimension of information, which can be seen as a potential 'alphabet' for its corresponding characteristic. At this point, the two-dimensional mapping matrices used to describe light field transformations can be expanded into high-dimensional tensor representations to more accurately handle the multi-degree-of-freedom information contained within the light field. In inverse design algorithms, the tensor slices corresponding to each DoF can systematically describe how the light field changes across specific DoFs. It is worth mentioning that our method provides a scalable path for realizing light field conversion within more DoFs.

Here, we have designed three types of inverse-designed components for valley topological photonic crystal which could manipulate phase, wavelength, spatial distribution. And they have potential to support future advanced topology quantum circuits with complex functional requirements. For example, Traditional topological phase shifters rely on path variations for phase delay, which highly dependent on wavelength. In Fig. 5a, we have developed a set of broadband phase shift devices based on inverse design, capable of introducing preset phase changes in topological photonic crystals. By integrating three R(θ) gates and two Hadamard gates, we can construct arbitrary unitary quantum gates within the topological photonic crystal framework (Supplementary Information section S7.1). Additionally, our design facilitates the creation of wavelength-division multiplexing (WDM) devices in these crystals, as shown in Fig. 5b. These devices, by achieving 'space-wavelength-valley' coherence, endow light with unique WDM functionalities as it passes through different valley states (IL<0.5 dB, ER>20 dB), which could be used for rerouting the entangle photons (Supplementary Information section 8). Our inverse design approach also allows spatial light to be converted into pseudo-spin modes, achieving direct coupling with topological photonic crystals, as shown in Fig. 5c. The device performance is comparable to traditional gratings, proving our ability to effectively control the spatial dimension of light and providing an interface to communicate quantum information between fiber and chips (Supplementary Information section 9).

**Conclusions**

By introducing a data-driven inverse design method, we have pioneered a paradigmatic shift in light manipulation, tapping into the latent 3D vectorial nature of on-chip structured light and offering a peerless degree of control within the light field. Incorporating our method into the topological photonics, which had attracted substantial research interest, a series of high-performance valley topological photonic crystal devices was demonstrated. These devices realized the conversion between arbitrary vectorial structured light in topological photonics, from standard waveguide mode to pseudo-spin mode, from topologically trivial interface state to topological non-trivial interface state, from single-mode topological interface state to multi-mode topological interface state, reducing the defects of high loss and fixed functions faced by traditional topological photonics, and promoting the practical application of topological photonics. Further, we extended the method to other DoFs, and design and implement a series of novel devices with high performance in topological optical networks, including quantum logic gate devices, wavelength division multiplexer, and free space couplers, which are expected to become the core devices of future large-scale on-chip quantum communication networks. The vector-based inverse design method provided a platform that is not only suitable for spin photonics but also capable of realizing arbitrary light field control on-chip, offering promising potential applications in photonic and quantum systems.

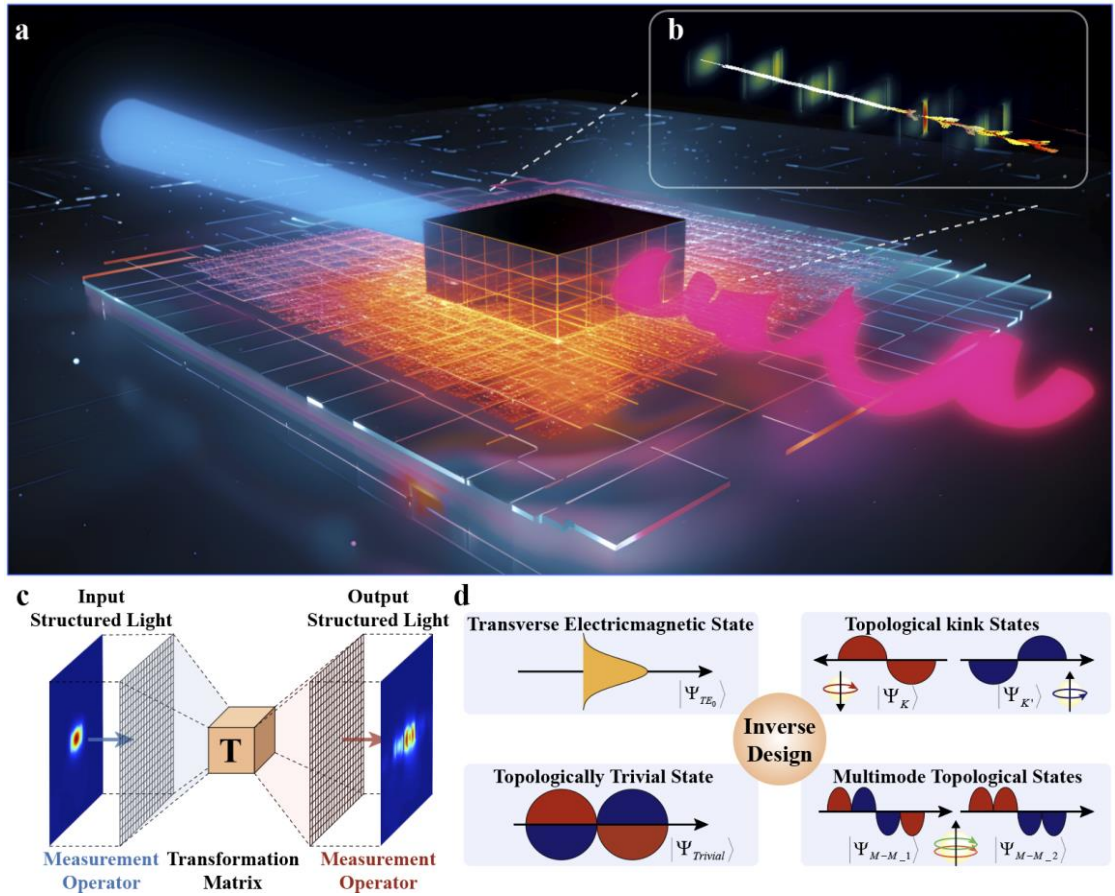

Fig.1. Inverse design to achieve arbitrary on-chip vectorial structure light conversion. (a) Schematic of the arbitrary on-chip structured light conversion. (Inset) Black area indicates the inverse-designed structure. (b) Light field evolution from TE0 mode to pseudo-spin mode. (Inset) Orange arrows represent wave vectors. (c) Flow chart of vector-based inverse design method. (d) Classification of structured light in topological photonics.

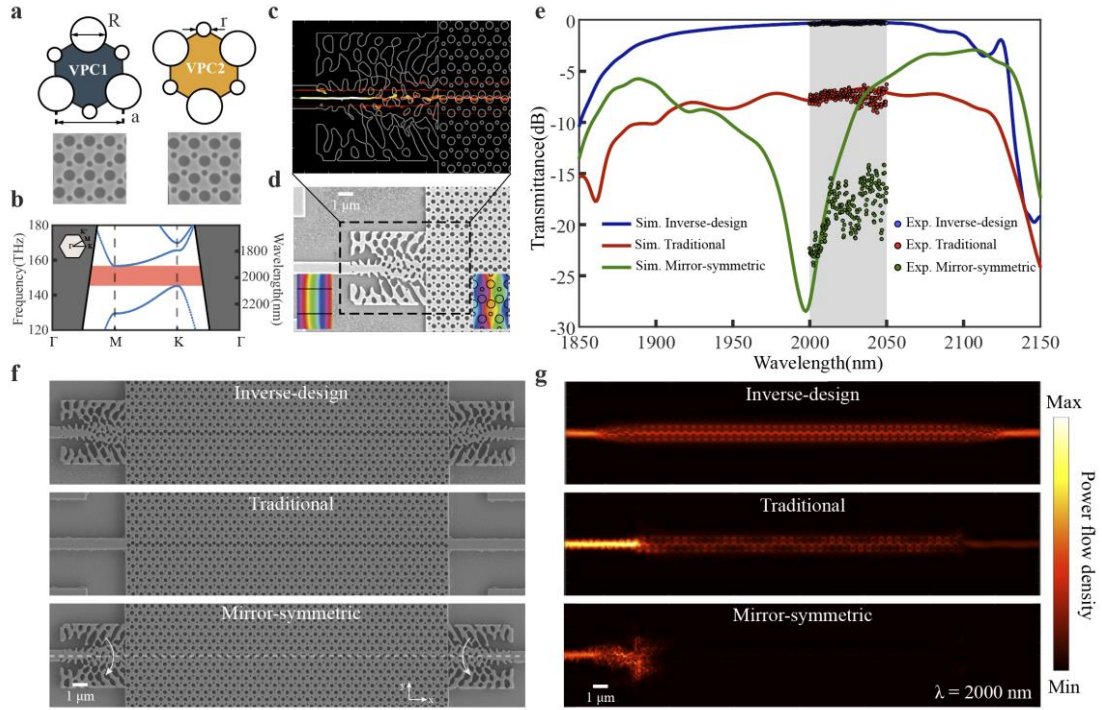

Fig.2. Mode conversion from TE$_0$ mode to pseudo-spin mode (VPC Coupler). (a) Details of two kinds of VPC unit cells (a=630 nm, R=296 nm, r=133 nm). (b) Bulk band both for VPC1 and VPC2. (c) Wave-vector distribution of VPC coupler based on inverse design. (d) Top-view SEM image of the sample. (Inset) H$_z$ phase of the TE$_0$ mode and the pseudo-spin mode. (e) Simulated (Line) and measured (dot) transmission spectra of different connections. Inverse design connection (Blue); Traditional single-mode waveguide connection (Red); Mirror-symmetric connection (Green). (f) Top-view SEM image of different connections. (g) Energy flow distribution of different connections at the wavelength of 2 μm.

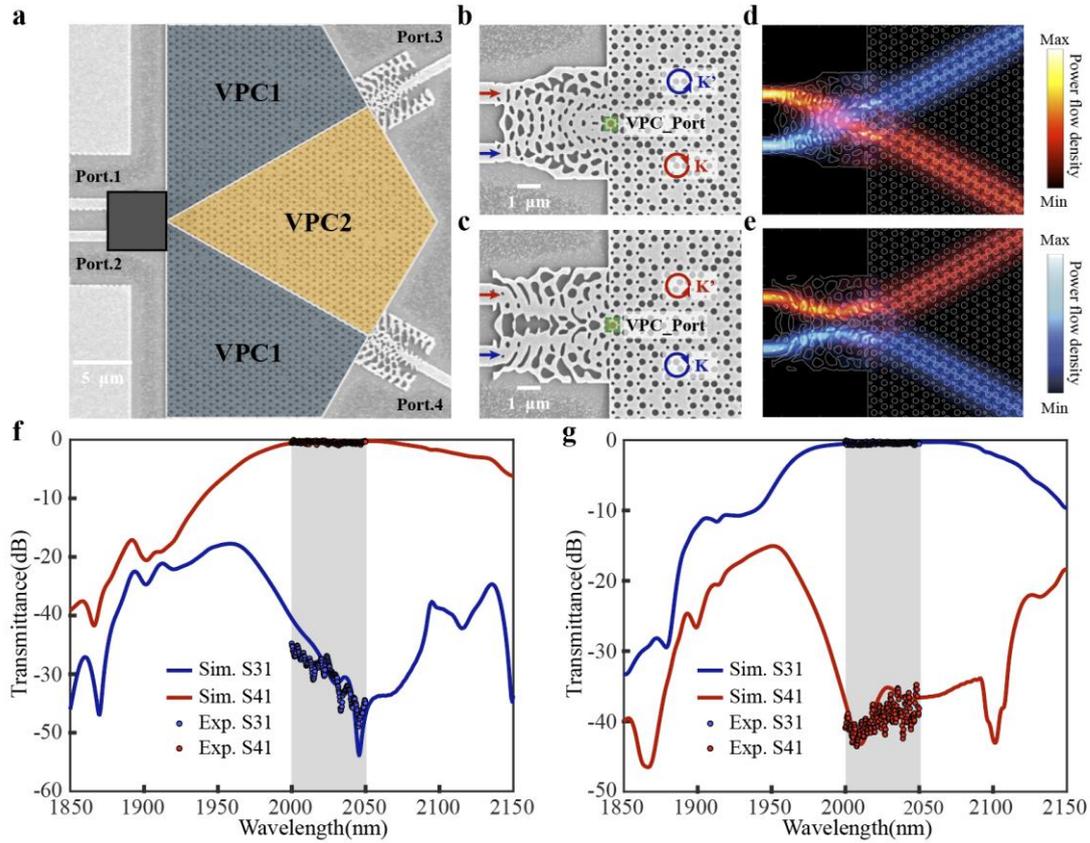

Fig.3. Mode conversion from TE$_0$ mode to arbitrary pseudo-spin mode (VPC router). (a) Top-view SEM image of the VPC router. (b and c) High-resolution top-view SEM image of the crossing route and the parallel route. (d and e) Energy flow distribution of the crossing route and the parallel route at the wavelength of 2 μm. (f and g) Simulated (Line) and measured (dot) transmission spectra of the crossing route and the parallel route.

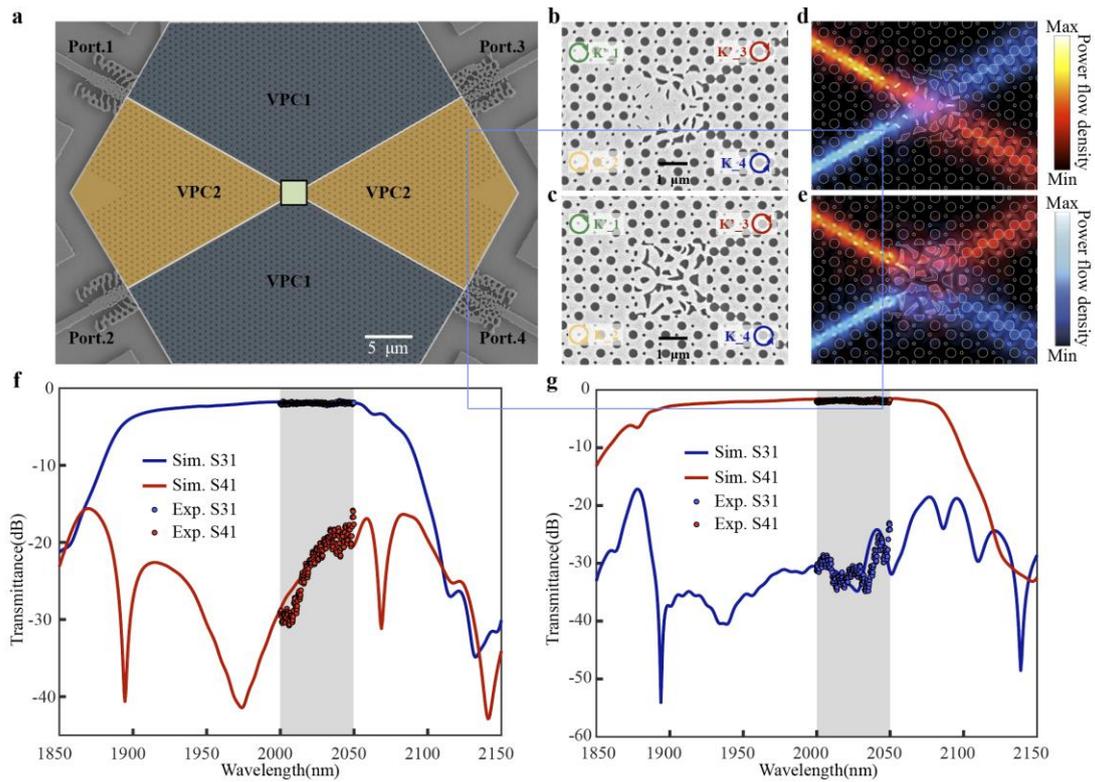

Fig.4. Mode conversion from arbitrary pseudo-spin mode to arbitrary pseudo-spin mode (VPC router). (a) Top-view SEM image of the VPC router. The structural parameters (a=650 nm, R=340 nm, and r=103 nm) was adjust to ensure that different pseudo-spin states have the same bandwidth. (b and c) High-resolution top-view SEM image of the crossing route and the parallel route. (d and e) Energy flow distribution of the crossing route and the parallel route at the wavelength of 2 μm. (f and g) Simulated (Line) and measured (dot) transmission spectra of the crossing route and the parallel route.

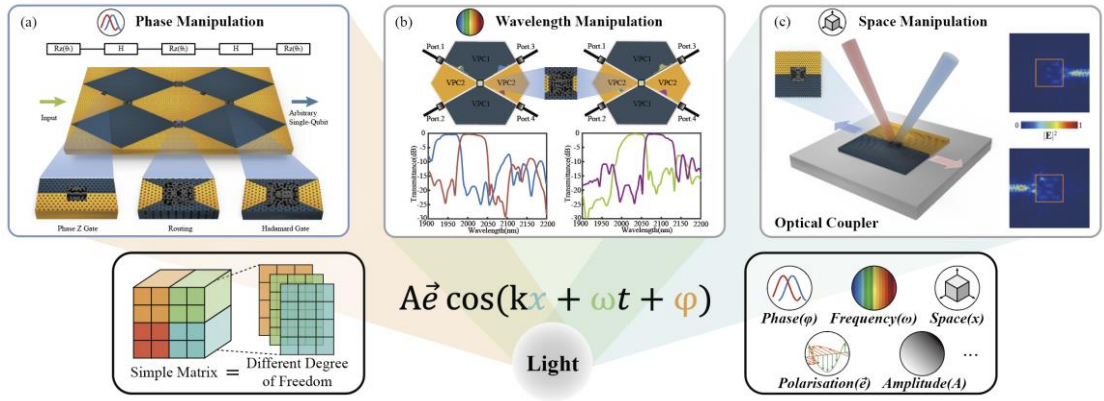

Fig.5. Manipulation of Other DoFs. (a) Phase manipulation for quantum logic gate design; (b) Wavelength manipulation for WDM device design; (c) Space manipulation for free-space coupler.

# Supplementary Materials for

## On-chip Vectorial Structured Light Manipulation via Inverse Design

Xiaobin Lin[1†], Maoliang Wei[1†], Kunhao Lei[1], Zijia Wang[1], Chi Wang[1], Hui Ma[1], Yuting Ye[2], Qiwei Zhan[1], Da Li[1], Shixun Dai[3], Baile Zhang[4], Xiaoyong Hu[5], Lan Li[2], Erping Li[1*], Hongtao Lin[1,6*]

[1]*State Key Laboratory of Brain-Machine Intelligence, Key Laboratory of Micro-Nano Electronics and Smart System of Zhejiang Province, College of Information Science and Electronic Engineering, Zhejiang University, Hangzhou 310027, China.*
[2]*Key Laboratory of 3D Micro/Nano Fabrication and Characterization of Zhejiang Province, School of Engineering, Westlake University, Hangzhou 310024, China*
[3]*Laboratory of Infrared Materials and Devices, Ningbo University, Ningbo 315211, China*
[4]*Division of Physics and Applied Physics, School of Physical and Mathematical Sciences, Nanyang Technological University, Singapore 637371, Singapore.*
[5]*State Key Laboratory for Mesoscopic Physics, Frontiers Science Center for Nano-optoelectronics, School of Physics, Peking University, Beijing 100871, China.*
[6]*MOE Frontier Science Center for Brain Science & Brain-Machine Integration, Zhejiang University, Hangzhou 310027, China.*

*\*Corresponding author. Email: hometown@zju.edu.cn, liep@zju.edu.cn*
†*These authors contributed equally to this work.*

**Methods**

Fabrication

The chalcogenide glass (ChG) film was composed of $Ge_{28}Sb_{12}Se_{60}$. The thin films (500 nm thickness) were deposited on top of a standard silicon wafer with 2 μm thick silicon oxide by thermal evaporation. The sample was exposed to UV light (100 mW cm$^{-2}$) for 3 days and baked for 3h (150 °C). The film was then spin-coated by the positive-tone e-beam resist (ARP 6200.13). The photonic crystals and the inverse design structure were patterned by using electron beam lithography (Raith VOYAGER). The ChG of the electron beam exposed area was then etched by inductively coupled plasma etching (Oxford Plasmapro100 Cobra 180) and the patterns were finally transferred to the film by the removal of the e-beam resist.

Optical measurements

The mid-infrared (mid-IR) test system for the measurement was mainly composed of a broadband tuning laser (Newfocus TLB-6700), a vertical fiber coupling platform, and a photodetector (PDA200C). The polarization direction of the light was controlled using a manual polarization controller. The transmission spectrum was measured by scanning the wavelength of the laser, and the output signal power was detected by the photodetector.

## Supplementary Text
## Section 1. On-chip structured light

The performance and functionalities of optical systems pivot around the scrupulous control of light's degrees of freedom (DoFs), such as spatial, temporal, and polarization characteristics. This has been further elucidated with the introduction of the concept of 'structured light', which significantly advanced our understanding and exploitation of these DoFs[1]. Specifically, structured light refers to the creation of particular light fields, achieved through meticulous manipulation and adjustment of these DoFs. The orbital angular momentum (OAM) is perhaps the most classical example of structured light, marked by its unique vortex wave propagation. As shown in Fig. S1a, this OAM type is delineated through topological charge numbers, presenting unbounded theoretical adaptations. For photons of circularly polarized light, be it left- or right-handed, they are attributed a spin angular momentum (SAM) of $+\hbar$ or $-\hbar$. For vortex light, each photon carries an OAM of $\pm\ell\hbar$, where "$\pm$" represents the chirality of the vortex light, and the range of topological charge $\ell$ is unlimited. Beyond this, structured light showcases configurations like the Möbius strips[2], functioning in the absence of OAM. The orchestration of optical Möbius strips (OMS) is anchored in modulating optical parameters in harmony with Möbius topological principles. The vortex light has been applied uncovers vast potential across fields, including optical communications[3], quantum entanglement[43], vortex optical tweezers[5], optical machining[6], and microscopic imaging[7].

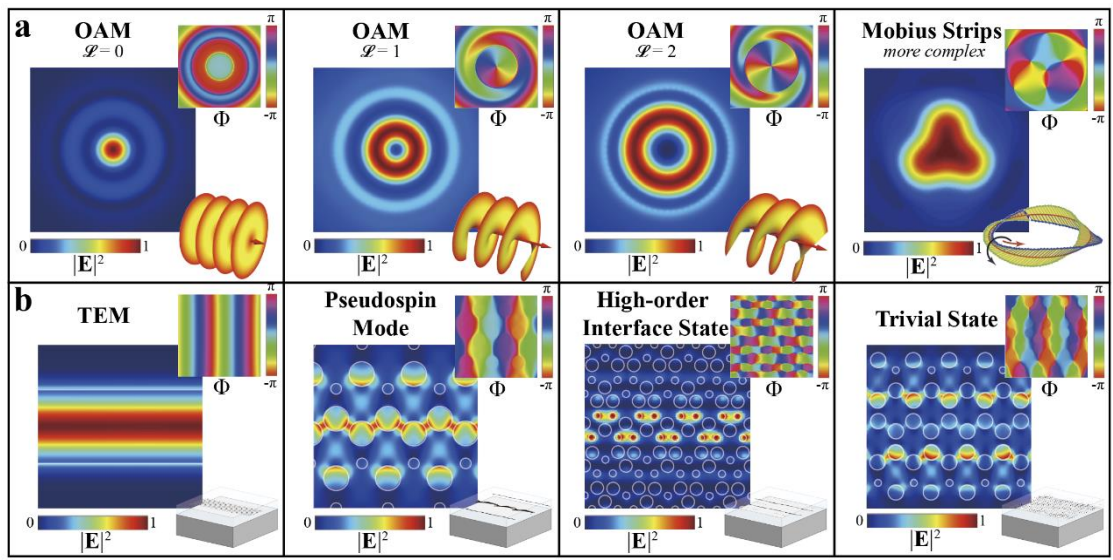

**Fig. S1. Structured light and on-chip structured light.** (**a**) Classic structured light exists in free space. (**b**) On-chip structured light exists in integrated photonics.

The development of structured light still faces many challenges, especially the transfer from bulk optics to integrated on-chip optics. In integrated optics, most research remains on standard transverse electromagnetic modes, which do not carry angular momentum information. Topological photonics is one of the most promising topics in recent years, and the topological degrees of freedom in photonics provide novel properties for unidirectional optical flow in various systems. In the valley

topology photonic crystal mentioned in the text, light propagates in the form of pseudospin[8]:

$$p_{\pm} = \frac{p_x \pm ip_y}{\sqrt{2}} \quad (S1)$$

where, $p_x$ and $p_y$ are the degenerate states of the Hamiltonian of photonic crystals, having the same parity as the Languet-Gaussian mode:

$$LG = \frac{HG_{10} \pm iHG_{01}}{\sqrt{2}} \quad (S2)$$

where, the $HG_{10}$ and $HG_{01}$ are the Hermite-Gaussian modes. At the same time, we can observe phase vortices in the center of the unit cell of the VPC, which also confirms the manuscript. Similar to the topological charge of OAM, the pattern of topological edge states also depends on the Chern number of the photonic crystal, as shown in Fig. S1b. In the Stampfli-triangle photonic crystal mentioned above, we construct a multi-mode topological photonic crystal waveguide. In the high-order mode, the phase vortex at the center of the unit cell is similar to the phase vortex of the OAM when $\ell=2$. Photonic crystals give us the possibility of three-dimensional control of on-chip light fields. Like the OMS, a complex structured light field may still exist in the photonic crystal without the Chern number difference. The analogy between on-chip structured light and structured light in free space has opened new ideas for the subsequent development of integrated optics.

## Section 2. Vector-Based Inverse Design Method
### S2.1. Data-Driven Inverse Design Optimization

In our research, the input and output structured light fields are predetermined, which in turn define the unitary operator $\hat{\mathbf{U}}$ and the mapping matrix $\mathbf{T}$. To facilitate the conversion of arbitrary on-chip structured light, as referenced in the main text, a structure within a predefined area that aligns with the corresponding mapping matrix is designed through our vectored-base inverse design method.

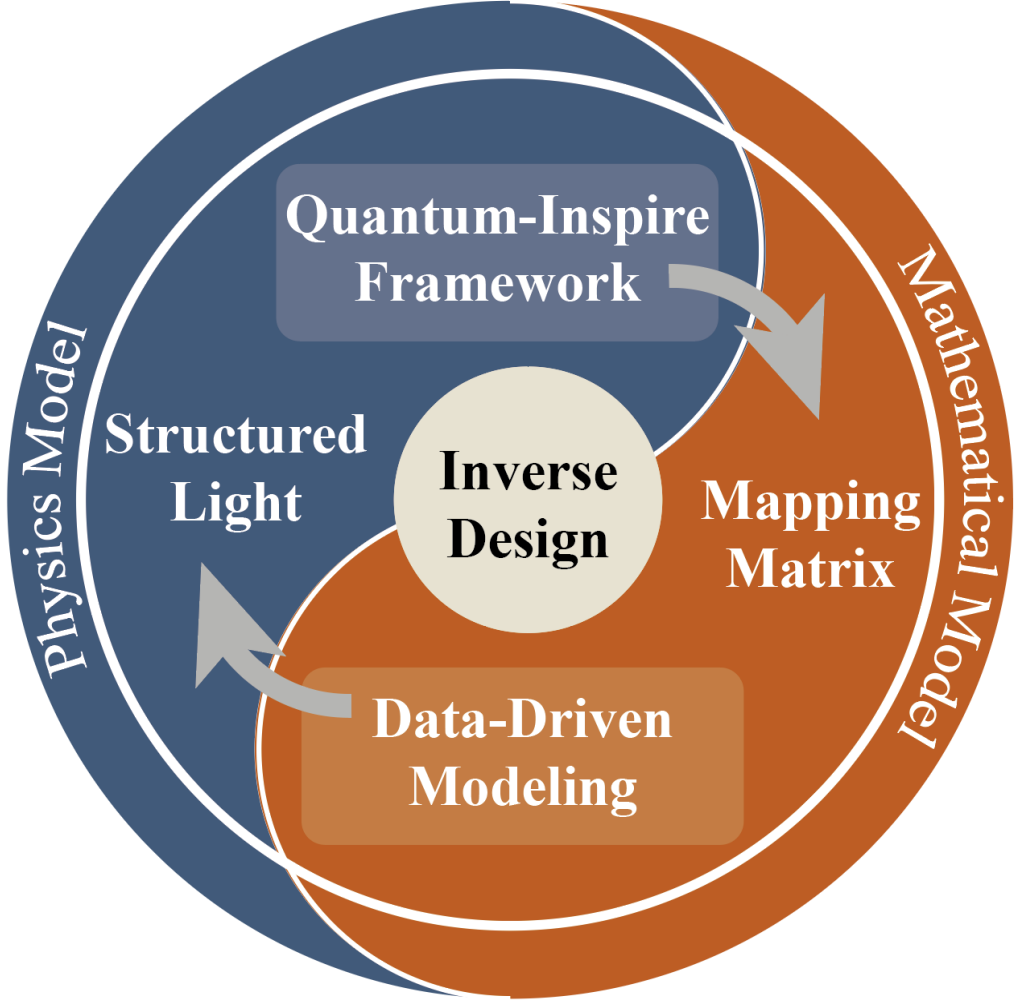

**Fig. S2. Schematic illustration of our inverse design algorithm.** A correlation is established between physical light fields and mathematical function spaces.

We've developed a comprehensive electromagnetic solver using a custom code, applying the discontinuous Galerkin method to divide the three-dimensional space into grids to solve Maxwell's equations. The idea of the objective-first method[9, 10] is borrowed here to achieve a fast solution to the optimized problem.

The system's electromagnetic problem can be expressed in a unified form, representing the physics residual:

$$\left(\nabla \times \mu_0^{-1} \nabla \times \omega^2 \varepsilon\right) E = i\omega J \rightarrow \mathbf{M} x = \mathbf{B} \tag{S3}$$

here, $x$ signifies the spatial vector electric field information, **M** is a sparse square matrix containing local permittivity and permeability information, and **B** represents the system's excitation source. Our inverse design problem can be viewed as the addition of an extra constraint, $\mathbf{T}x = \mathbf{C}$, to the passive system ($\mathbf{M}x = 0$), ensuring alignment with the mapping matrix **T**. We use the least squares minimization to reduce the residual error in the field to meet Maxwell's equations.

To further obtain the corresponding structure permittivity distribution, we calculate the spatial permittivity distribution by optimizing the adjoint electromagnetic problem, with the changes in permittivity represented as gradients. With continuous iterations, we obtain the optimal permittivity of the inverse-design structure. As shown in Fig. S2, we learn physic-based mode from data and provide a paradigm to realize arbitrary on-chip structured light manipulation. By introducing a mapping operator linking arbitrary structured light fields under a quantum-inspired framework, we establish a correlation between physical light fields and mathematical function spaces.

## S2.2. Robust-Inverse Design Strategy

Relative to conventional design methods, the inverse design methodology demonstrates superior dimensionality, enabling the fabrication of devices endowed with intricate functionalities and compact dimensions. However, this method's penchant for minuscule feature sizes renders the resultant devices particularly vulnerable to manufacturing discrepancies. Inherent random variabilities – stemming from material composition inconsistencies, geometric discrepancies, and fabrication missteps – invariably impact the device's performance. Such susceptibilities to manufacturing imperfections have been a notable bottleneck for the deployment of inverse-designed photonic devices.

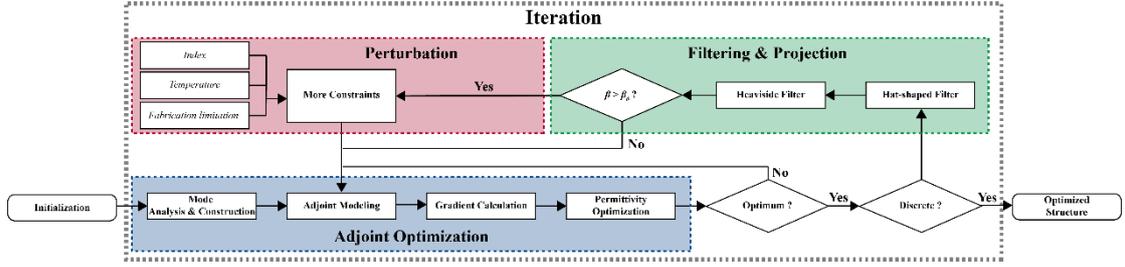

**Fig. S3. Schematic illustration of the robust-inverse design strategy.** The detailed optimization process of the design algorithm: "Adjoint Optimization", "Filter & Projection" and "Perturbation".

Here, we proposed a new robust inverse design strategy. This enhanced adjoint method unfolds in a tripartite sequence: "Adjoint Optimization", "Filtering & Projection", and "Perturbation", as illustrated in Fig. S3. Ensuring the relative permittivity continuity between air and ChG, the gradient concerning material density is ascertained via the adjoint sensitivity analysis. Through iterative refinement, the device's objective function converges to an optimal value. Subsequently, we employ a three-field method[1111] to transition from continuous material density to distinct values:

$$\rho_i = \frac{\sum_{j \in N_i} \omega(X_j) v_j \rho_j}{\sum_{j \in N_i} \omega(X_j) v_j}, \quad \omega(X_j) = R - |X_i - X_j| \tag{S4}$$

Firstly, we use a linear hat-shape filter to eliminate small feature structures. where $\rho_i$ is the design field parameter and is the filtered field parameter. $N_i$ is the neighborhood set of elements lying with the filter domain of the element $i$. $R$ is the radius of a linear hat-shape filter. $v_j$ is the volume of the element $j$, $\omega$ is a weighting function of the distance between the central coordinates $X_i$ and $X_j$ of the cell $i$ and $j$. Secondly, we used a Heaviside filter to binarize the optimized structure:

$$\overline{\rho_i} = \frac{\tanh(\beta \cdot \eta) + \tanh(\beta \cdot (\rho_i - \eta))}{\tanh(\beta \cdot \eta) + \tanh(\beta \cdot (1-\eta))} \tag{S5}$$

where $\eta$ is the threshold, and $\beta$ is the binarization factor, which controls the steepness of the Heaviside function. By using the three-field method, the design field parameter $\rho_i$ was projected to the physical field parameter $\overline{\rho_i}$, and the design structure was able to be fabricated. By projecting the design field onto the physical field, we obtained the optimized physical structure of the area with no intermediate states.

Thus far, our design has strictly adhered to physical parameters. Yet, manufacturing uncertainties, which are inherent and random, potentially threaten to skew performance metrics and complicate the modeling. In the "Perturbation", we incorporated perturbations reflective of manufacturing discrepancies into our optimization workflow and proposed a novel robust inverse optimization strategy:

$$\min_{\rho} \; F(\rho) = \begin{cases} F_0(\rho) & \beta \leq \beta_0 \\ F_0(\rho) + E_{\xi}\left[\left(F^*(\rho,\xi) - F_0(\rho)\right)^2\right] & \beta > \beta_0 \end{cases} \quad \text{(S6)}$$

where $\beta$ was the binarization factor of the Heaviside filter in the three-field method, and $\beta_0$ was the threshold value that we set. $\xi$ was the vector describing fabrication variations, such as material composition fluctuations, geometry deviations, and processing errors. $F^*(\rho, \xi)$ was the target under perturbation, which described the influence of manufacturing variations.

When evaluating the impact of manufacturing uncertainties on the functionality of optical systems, we adopt the Generalized Polynomial Chaos (gPC) expansion method[12]. The core idea of gPC is to approximate the output of interest using a set of orthogonal polynomial basis functions. Here, the function is expanded using gPC:

$$F^*(\theta,\xi) \approx \sum_i C_i(\theta)\Psi_i(\xi_i) \quad \text{(S7)}$$

where $\Psi_i(\xi_i)$ is the multivariate polynomial corresponding to the $i$th error term, $C_i(\theta)$ is the corresponding coefficient, $i$ is a set of finite-dimensional non-negative integers, and $i \in (i_1, i_2, \cdots i_d)$. At this point, $i$ can be mapped to an array from 1 to $N$:

$$\begin{aligned} i_1 + i_2 + \cdots + i_d &\leq p \\ N &= \frac{(d+p)!}{p!d!} \end{aligned} \quad \text{(S8)}$$

and the function can be rewrite:

$$F^*(\theta, \; \xi) \approx \sum_{n=1}^{N} c_n(\theta)\psi_n(\xi) \quad \text{(S9)}$$

In gPC expansion, different probability distributions correspond to different orthogonal polynomial bases. Considering that the error probability density function conforms to a uniform distribution, the formula can be expanded and written as:

$$F^*(\theta,\xi) \approx \sum_{a,b} C_{a,b} L_a(\theta)\Phi_b(\xi) \quad \text{(S10)}$$

where **L** is a multivariate Legendre polynomial, $\Phi$ is the corresponding multivariate polynomial.

Then we can get:

$$E_\xi\left[F^*(\theta,\xi)\right] = \int F^*(\theta,\xi)\rho_\xi d\xi$$
$$= \sum_{a,b} C_{a,b} L_a(\theta) \int \Phi_b(\xi)\rho_\xi d\xi \quad (S11)$$

In the gPC model, the polynomials used (such as Legendre polynomials) are orthogonal. Orthogonality means that the inner product between different polynomials under the corresponding weight functions is zero. Also, the value of the lowest order (usually zeroth order) of these polynomials is a constant (e.g. 1). Therefore, when using these properties to perform mathematical operations or expansions, calculations can be simplified based on orthogonality to derive specific conclusions or formulas:

$$E_\xi\left[F^*(\theta,\xi)\right] = \sum_a C_{a,\vec{0}} L_a(\theta) \quad (S12)$$

so, the object function can be write as:

$$F(\theta,\xi) = F_0(\theta) + V_\xi\left(F^*(\theta,\xi)\right) + \left(E_\xi\left(F^*(\theta,\xi)\right) - F_0(\theta)\right)^2$$
$$= \left(F_0(\theta) - 2\sum_a C_{a,\vec{0}} L_a(\theta)\right) F_0(\theta) + F_0(\theta) +$$
$$\sum_{a,a',b\neq\vec{0}} C_{a,b} C_{a',b} L_a(\theta) L_{a'}(\theta) + \left(\sum_a C_{a,\vec{0}} L_a(\theta)\right)^2$$
$$= F_{NEW}(\theta) \quad (S13)$$

we brought the $F_{NEW}(\theta)$ into the adjoint method, and the gradient of the device objective function with respect to the optimization parameters under error disturbance can be calculated at one time. In addition to the uniform distribution, in the gPC expansion, each probability density function has a corresponding set of orthogonal polynomials. When the Legendre polynomial and coefficients are known, the amount of calculation of the derivatives will increase, but this increase is limited.

During the manufacturing process, manufacturing variations are randomly distributed and independent of each other. Given the probability density function (pdf) of $\xi$, we introduced the expectation of the mean square error $E_\xi[(FOM^*(\rho,\xi)-FOM(\rho))^2]$ into the objective function, in order to enhance the robustness of the design.

## Section 3. Abnormal Transmission of the Mirror-Symmetric Inverse Design Structure

In the VPC, light primarily traveled in the form of a trivial state at the edge of the band[13, 14], as illustrated in Fig. S4a. The most striking difference between trivial and topological states was that when we excited an K'-valley mode in the topological waveguide, the light propagated in a single direction in the topological state. Whereas in the trivial state, the light could propagate in both directions (Fig. S4d to f). From the band diagram, we observed that the slope of the K'-valley mode band had both positive and negative values in the trivial state, while in the topological state, the band slope only had positive values.

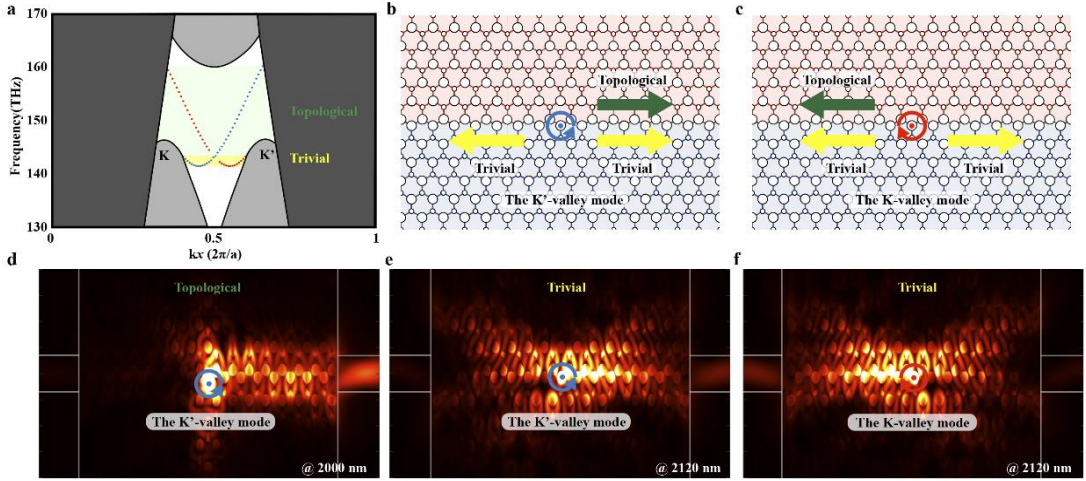

**Fig. S4. Trivial states in the VPC.** (**a**) The projected band structure of the VPC waveguide. (**b** and **c**) Propagation of different pseudospin modes in the VPC waveguide. (**d** to **f**) The energy flow distributions.

In the main text, we discovered that the inverse-designed device exhibited abnormal transmission at the edge of the band when mirror symmetry was applied. This was because the light entering the device transitioned from the K'-valley mode to the K-valley mode due to mirror symmetry. In the trivial states, the device also supported efficient transmission of the K-valley mode (Fig. S5a and b). This indirectly confirmed the high conversion efficiency of our device for the specified pseudo-spin mode.

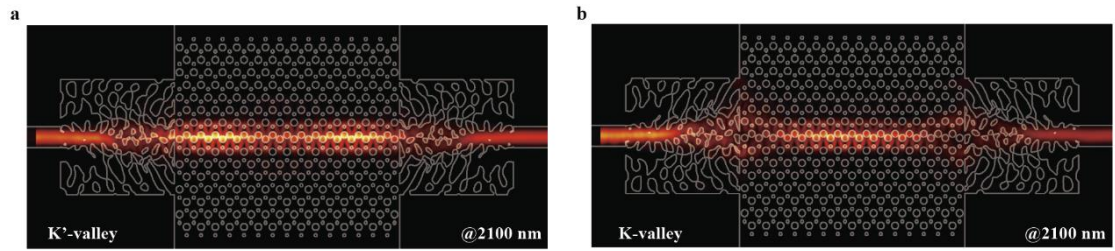

**Fig. S5. The abnormal transmission at the trivial state. a,** The energy flow distribution of the device when the K'-valley mode was input (inverse design connection). **b,** The energy flow distribution of the device when the K-valley mode was input (mirror-symmetric connection).

## Section 4. The Eye Diagram of Inverse Design Devices

We conducted digital signal test on the inverse design devices, and the test eye diagram is shown in Fig. S6. Using a continuous wave (CW) laser as the input source, the light is modulated by a signal generator. The output light is amplified by a praseodymium doped fiber amplifier (PDFA) to compensate for link losses. 4.25 Gb/s is the upper limit of the current experiment equipment, and higher data rate testing can be achieved with higher bandwidth oscilloscopes and signal generators. We found that devices based on reverse design, from couplers to routers, can achieve eye diagram results similar to those of traditional single-mode waveguides.

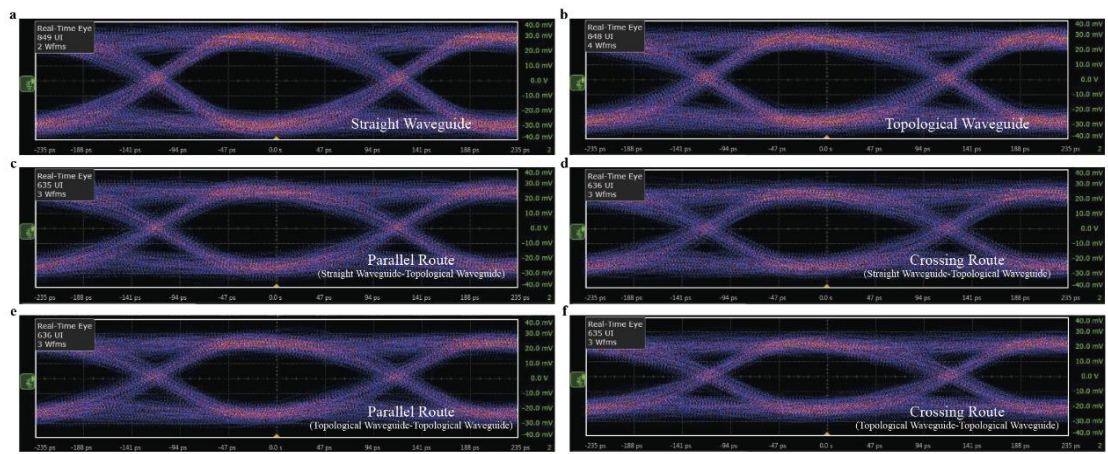

**Fig. S6. The eye diagram of inverse design devices.** (**a**) The straight waveguide. (**b**) The topological waveguide with inverse design coupler. (**c**) The inverse design parallel route (straight waveguide-topological waveguide). (**d**) The inverse design crossing route (straight waveguide-topological waveguide). (**e**) The inverse design parallel route (topological waveguide-topological waveguide). (**f**) The inverse design crossing route (topological waveguide-topological waveguide).

## Section 5. Topological Trivial State Coupler based on Inverse Design

Based on the previous topological waveguides, we changed the additional two rows of holes near the interface from small holes to large ones, as shown in Fig. S7b and c. We simulated the band of the corresponding multi-hole topological waveguide, as shown in Fig. S7a. In addition to the topological states of the topological waveguide, a new band at a low-frequency position was found. The multi-hole topological waveguide can be considered a defect to the traditional VPC waveguide, and the mode count of the topological states depends on the difference in valley Chern numbers between the two structures on either side. The additional band is not a topological state, but a trivial state.

By directly connecting the standard single-mode waveguide and the multi-hole topological waveguide, we simulated the transmission of "Flat" type and "Z" type topological waveguides. As shown in Fig. S7d, the device achieved signal transmission at 1875~2035 nm, with IL of 7.29 dB and 8.96 dB, respectively. In the range of 2183-2241 nm, the transmission was only achieved in the "Flat" type topological waveguide (with an IL of 7.22dB), but the sharp-angle transmission was not realized (Fig. S8a).

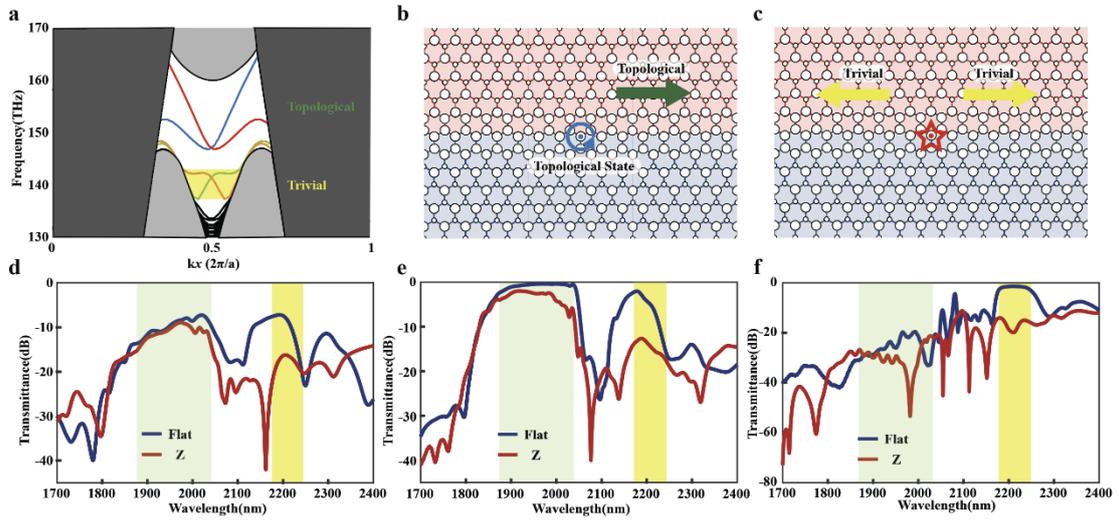

**Fig.S7. Multi-Hole VPC Waveguide.** (**a**) The projected band structure of the multi-hole VPC waveguide. (**b** and **c**) Propagation inside topological photonic crystals in different states. Simulated transmission spectra of different connections: Traditional single-mode waveguide connection (**d**) Inverse design structure in the main text **e**; Inverse design structure for the trivial state **f**.

We noticed that in the multi-hole topological waveguide, the band structure was similar to that of the traditional topological waveguide, and the mode also propagated in the form of pseudospin. We connected the inverse-design coupler in the main text, which was designed for the traditional topological waveguide, with the multi-hole topological waveguide and tested the transmission performance of the device (Fig. S7e). We found that in the topological state, the inverse-designed structure could effectively reduce the IL of the device (IL=0.38 dB), proving that our design is also suitable for topological waveguides with defects. However, in the trivial state, the insertion loss of the inverse-designed device was 2.13 dB, due to the device transmitting in a pseudospin-like mode at that time (Fig. S8b).

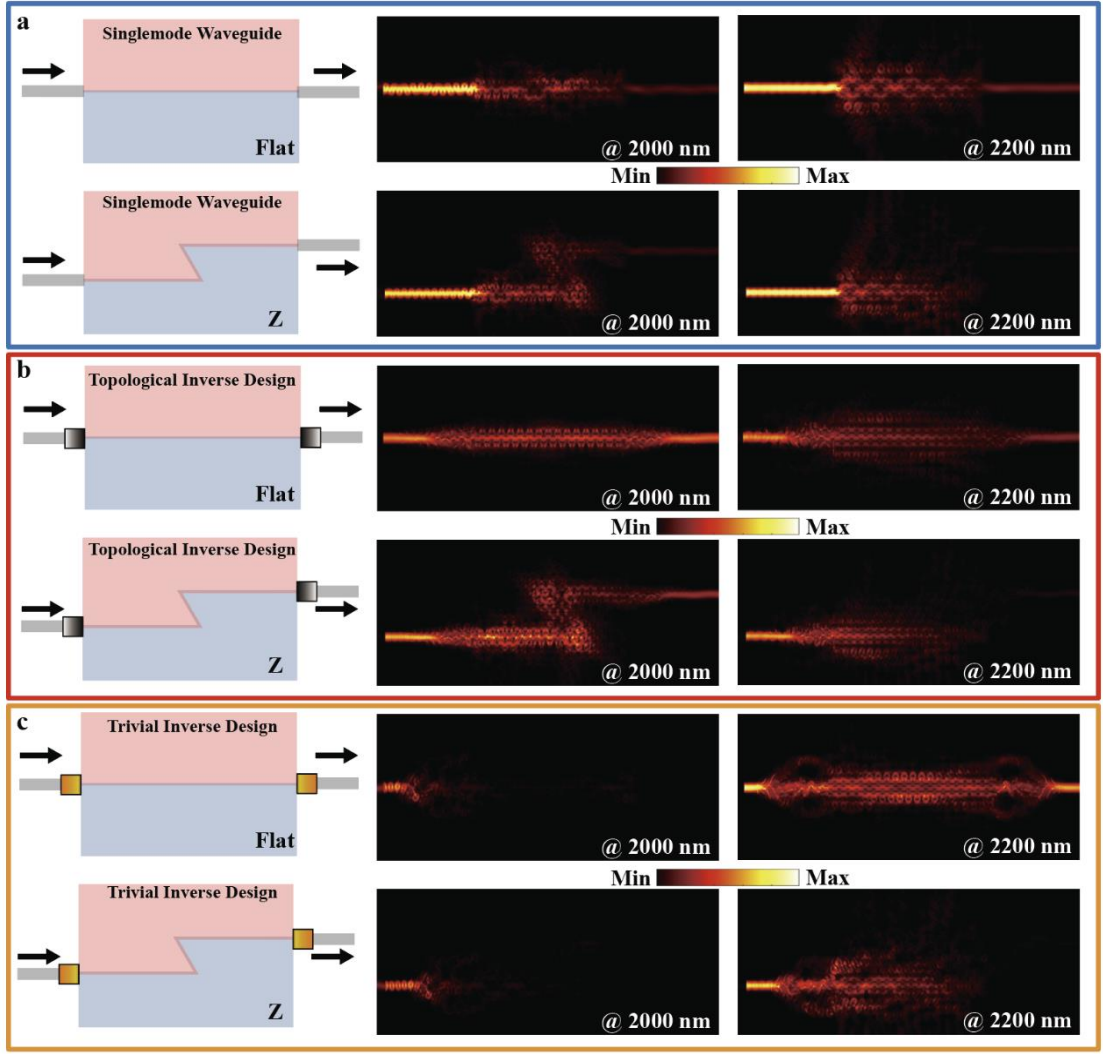

**Fig.S8. Propagation of multi-hole VPC waveguides ("Flat" and "Z" type) in different states under different connections.** (**a**) Traditional single-mode waveguide connection. (**b**) Inverse design structure in the main text. (**c**) Inverse design structure for the trivial state.

Further, by rewriting the objective function, we optimized for the bulk state separately:

$$\mathbf{E}_{Trivial} = \mathbf{T} \cdot \mathbf{E}_{TE_0} \tag{S14}$$

As shown in Fig. S8f, the device achieved efficient transmission in the trivial state with an IL of 1.37 dB ("Flat" type). In the topological state, the device achieved a maximum extinction ratio (ER) of 53.33 dB ("Z" type). The energy flow distribution of the corresponding wavelengths in the device can be observed in Fig. S8c. At 2000 nm, regardless of the "Flat" or "Z" type topological waveguide device, the energy could not be coupled into the topological waveguide and was mainly reflected. At 2200 nm, for the "Flat" type, efficient transmission of energy was achieved, while for the "Z" type, although the light was coupled into the topological waveguide, it mainly diverged within the photonic crystal instead of undergoing sharp transmission. Our inverse design method is not only applicable to pseudospin modes in the topological state but also suitable for efficient conversion in the trivial state, truly enabling the control of arbitrary mode fields.

## Section 6. Multimode Topological Photonics Crystal Coupler based on Inverse Design

Valley topological photonic crystal (VPC) devices have propelled advancements in communication[13], lasers[15], and quantum computation[16]. However, most existing devices operated solely at a single frequency and a single mode, presenting limitations to their potential applications. In topological photonic crystals, the number of modes in topological waveguides depended on the difference in topological invariants (valley Chern numbers) between the two structures at the interface[17, 18]:

$$C_n = \frac{1}{2\pi} \int_{BZ} \Omega_n(k) d^2k$$
$$\Omega_n(k) = \nabla \times \langle u_n(k) | i\nabla k | u_n(k) \rangle \quad (S15)$$

The valley Chern number ($C_n$) was a topological invariant obtained by integrating the Berry curvature in a closed Brillouin zone and the Berry curvature ($\Omega_n(k)$) was expressed by the sum of eigenstates. Here, $u_n(k)$ was the eigenstate of the $n$ band. When the system has time-reversal symmetry, the Berry curvature was an odd function about the wave vector $k$: $\Omega_n(-k) = -\Omega_n(k)$; while for the system with space-reversal symmetry, the Berry curvature was about the wave vector Even function of $k$: $\Omega_n(-k) = \Omega_n(k)$. In the VPC, we can break the spatial symmetry of the structure by changing the size of the holes/cylinders and achieve non-zero Chern numbers. The symmetry of the structure was reduced from $C_6$ symmetry to $C_3$ symmetry. The degeneracy at the Dirac point was broken, and the distribution of the Berry curvature at the K point and K' point is opposite.

Higher symmetry and rich energy band structure make the photonic quasicrystals considered to have greater potential to realize large Chen number VPC[14]. Here, a Stampfli-triangle photonic crystal[19] was proposed, which could be used to realize multi-mode, multi-frequency VPC. The schematic diagram of the structure is shown in Fig. S8a. The lattice constant a=1 μm, and the distance between the adjacent cylinders b=($\sqrt{3}$-1)/2 μm. We broke the inversion symmetry of the VPC (R=256 nm, r=183 nm) to open the bandgap between 1780 and 1905 nm, as shown in Fig. 2b in the main text.

As depicted in Fig. S9c, two multi-mode VPCs formed a "zigzag" type topological waveguide. Within the wavelength range of 1780~1905nm, the difference in Chern numbers at the K/K' point for the two topological photonic crystals was calculated to be 2 ($\left| C_{K/K'}^{Multi\_VPC1} - C_{K/K'}^{Multi\_VPC2} \right| = 2$). Fig. S9d displayed the simulated band structure of the topological waveguide, validating that the number of modes for the topological edge state was 2. The distribution of the $E_z$ electric field of the corresponding eigenmode for these edge states was presented in Fig. S9e.

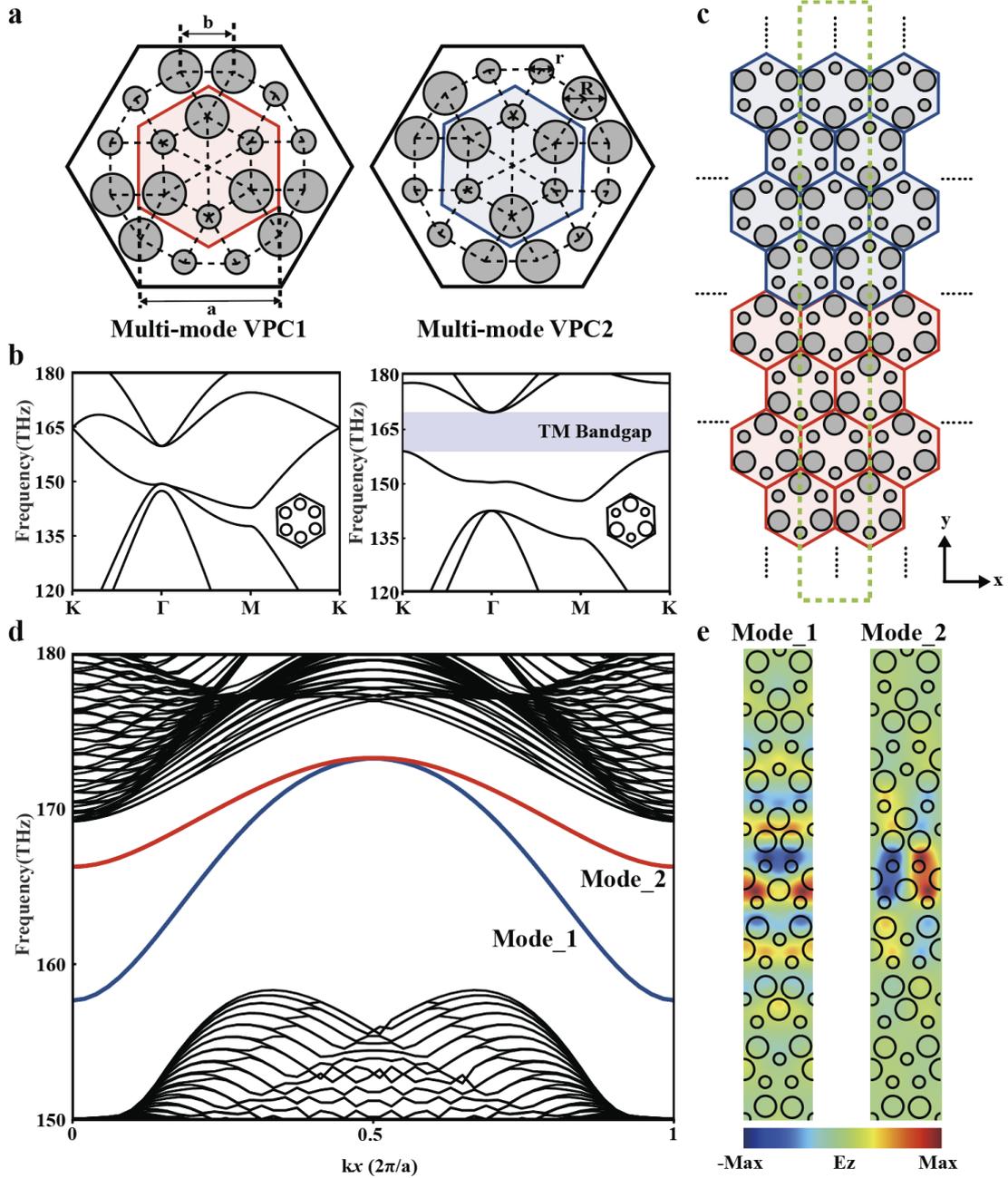

**Fig. S9. Multimode Topological Photonics Crystal.** (**a**) Details of two kinds of Muti-mode VPC unit cells (a=1 μm, b= (√3-1)/2 μm, R=256 nm, r=183 nm). (**b**) Bulk band for the muti-mode VPC. (**c**) The supercell of the multi-mode topological waveguide ("zigzag"). (**d**) The projected band structure. (**e**) The electric field distributions of the eigenmodes.

In order to verify that we achieved a multimode topological edge state, we simulated the transmission of a "Z" type topological waveguide when connected to a single-mode straight waveguide (Fig. S10A). The device achieved acute angle transmission within the operating band, with an insertion loss (IL) of 7.01 dB. The spectral lines generally exhibited oscillation, which was caused by a mode mismatch between the straight waveguide and the topological waveguide. Simultaneously, by analyzing the band structure of the topological waveguide (Fig. S9d), we noted that when the wavelength was between 1780~1807 nm (region I), the topological waveguide supported two edge state modes simultaneously. As the wavelength increased (1807-1905 nm, region II), the topological waveguide only supported a single mode. The mode transmission of the device is shown in Fig. S10b. We simulated the electric field distribution of the device at corresponding wavelengths. At 1850 nm, the light was transmitted in a single mode, and the electric field was consistent with mode_1 in Fig. S9e. However, at 1800 nm, the electric field was transmitted in a mixture of mode_1 and mode_2. For both waveguides, we observed strong reflection at the incident waveguide due to mode mismatch.

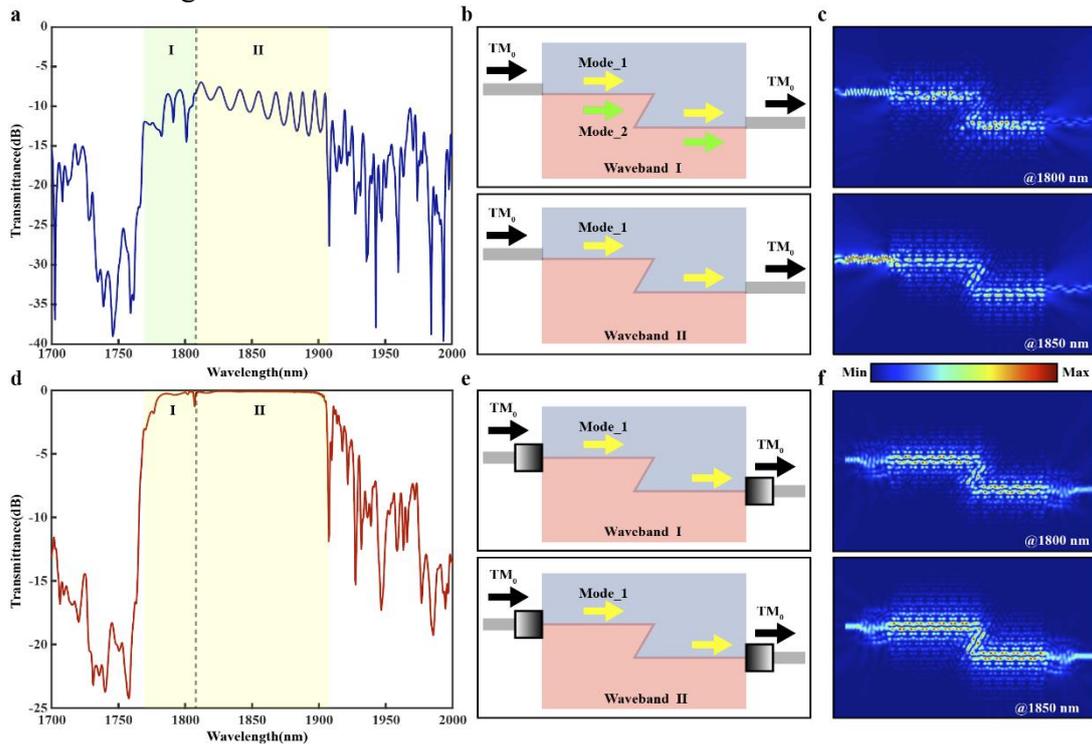

Fig. S10. Mode conversion from TM$_0$ mode to a specified mode in a multi-mode VPC (multi-mode VPC coupler). (**a**) Simulated transmission spectra of traditional single-mode waveguide connection; (**b**) Schematic diagram of mode transmission of devices in different wavebands when the single-mode waveguide was connected. (**c**) The electric field distributions of the single-mode waveguide connections at the wavelength of 1800 nm and 1850 nm. (**d**) Simulated transmission spectra of inverse design structure connection; (**e**) Schematic diagram of mode transmission of devices in different wavebands when the inverse design structure was connected. (**f**) The electric field distributions of the inverse design structure connections at the wavelength of 1800 nm and 1850 nm.

High IL and mixed modes have severely hindered the development of multimode topological photonic crystal devices. By revising the objective function which we mentioned in this work, we successfully achieved efficient conversion from the standard silicon waveguide TM$_0$ mode to the specified mode in the multimode topological waveguide (Fig. S10e):

$$\mathbf{E}_{Multimode\_1} = \mathbf{T} \cdot \mathbf{E}_{TE_0} \tag{S16}$$

where, $\vec{c}_{mode\_1}$ was the electric field distribution of the mode_1. We simulated the performance of the inverse-designed device, as shown in Fig. S10d. The device achieved efficient transmission throughout the entire operating waveband, with a low IL of 0.04 dB. To verify the purity of the mode, we simulated the electric field distribution of the device at corresponding wavelengths. At 1800nm and 1850nm, the light was transmitted in the same form as mode one, confirming that our device effectively converts to the specified mode in the topological edge state. This achievement was also a first in the realm of such devices.

## Section 7. Quantum Optical Circuit Design

Achieving a universal quantum computer is a significant goal in the field of quantum information science. Cascading a series of basic quantum gates enables the construction of quantum computers capable of performing arbitrary computational tasks. Quantum photonic integrated circuits, in particular, are considered a promising approach for large-scale quantum information processing by cascading various types of quantum gates to design and build qubits and quantum operations[20, 21]. Using multi-layer Mach-Zehnder Interferometers (MZIs), researchers have successfully created quantum circuits capable of qubit operations[22].

Valley topological photonic circuits have unique electromagnetic properties that make them significantly stable and resistant to external noise and interference, which are distinct from traditional optical systems. These properties are especially beneficial in photonics applications, notably in quantum information processing, where they contribute to the development of stable qubits and quantum gates[16, 23]. However, the functions of valley topological photonic circuits are currently limited due to the complexity of their modes, which present substantial challenges in terms of control and manipulation.

The Fig. S11a illustrates a quantum logic gate circuit engineered through inverse design, operating inside a topological photonic crystal. As laser pulses propagate within the topological photonic crystal, it efficiently generated correlated signal and idler photons across a broad bandwidth via spontaneous four-wave mixing[16, 23]. The produced signal and idler photons exhibit quantum entanglement. Utilizing a inverse design wavelength division multiplexing (WDM) device, the entangled photons are allocated to different topological edge states and transmitted to distinct quantum logic gates. Through a series of quantum logic gate devices, encoding of the entangled photons is achieved, as shown in Fig. S11b. Finally, the photons are coupled out of the chip through a free-space coupler (FSC), ready for subsequent communication or computation tasks.

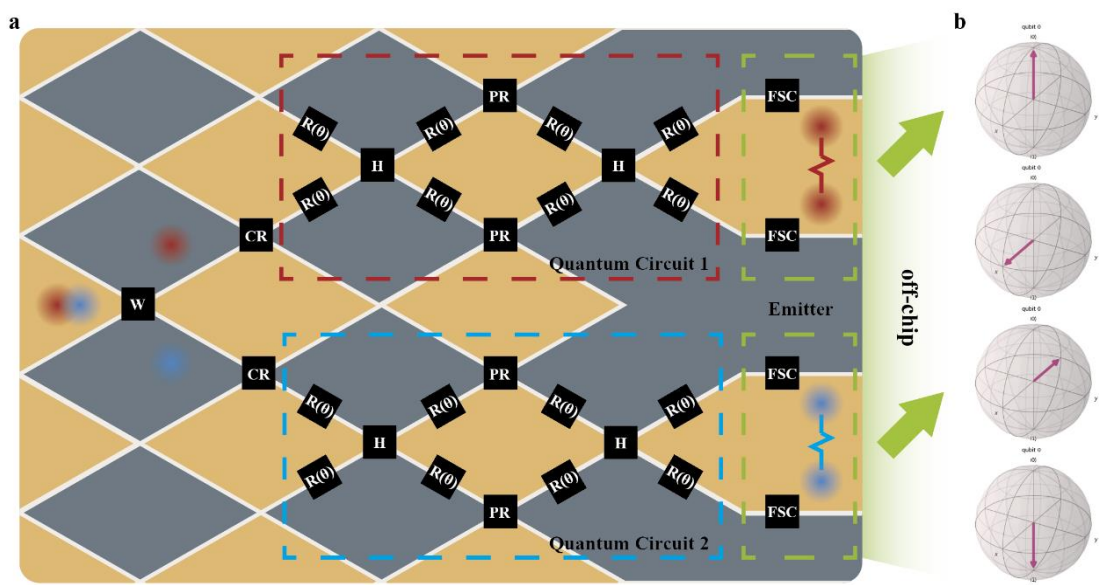

**Fig. S11. Quantum Optical Circuit Design.** (a) Schematic diagram of the quantum

logic gate circuit operating within a topological photonic crystal. (**b**) Different qubits generated based on quantum logic gates.

### S7.1. Quantum Logic Gate

Here, we utilize our method to construct a phase rotation gate. The Rz(θ) gate is easily implemented by introducing a phase difference between two paths of a photonic quantum state, creating a phase shift between the quantum states in the topological waveguide, as illustrated in Fig. S12a. Fig. S12b, c, and d display the Hz electric field distribution of the device at θ=0, π/2, and π, respectively, demonstrating precise phase variations.

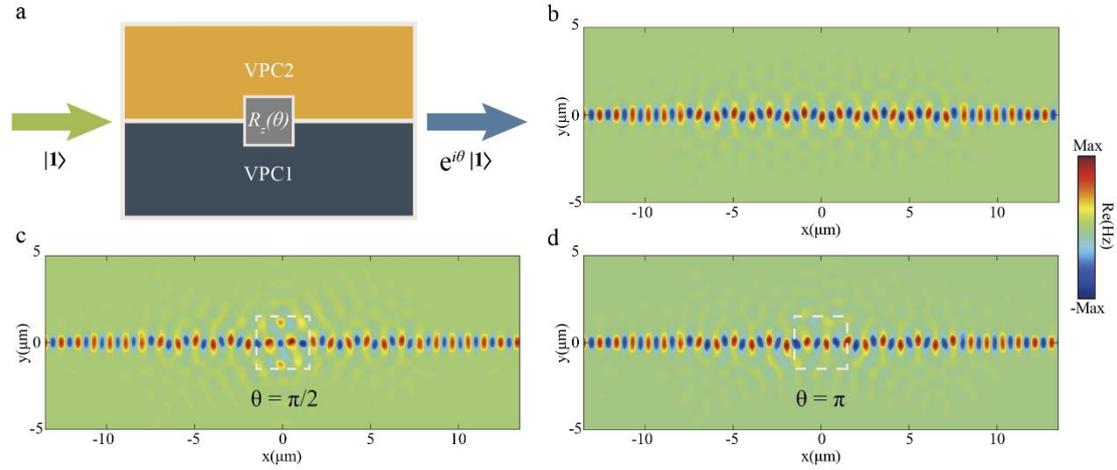

**Fig. S12. Phase Rotation Gate.** (**a**) Inverse design phase Rz(θ) gate. Simulation result of the phase z gate when θ = 0(**b**), π/2(**c**), π(**d**).

Based on the principle of single-photon interference, a Hadamard operation can be executed within the designed structure. When a single-photon state is injected into waveguide $a_{in}$ or $b_{in}$, it is transformed into a superposition state at the output aout or bout with a π/2 phase difference after passing through the Hadamard gate, as shown in Fig. S13a and b. Fig S13c and d show the simulated results of the field distribution under single-photon excitation at $a_{in}$ or $b_{in}$, while Fig. S13e and f confirm that the inverse-designed structure exhibits the good performance of a low-loss 50:50 beamsplitter with a stable phase difference across the operating bandwidth.

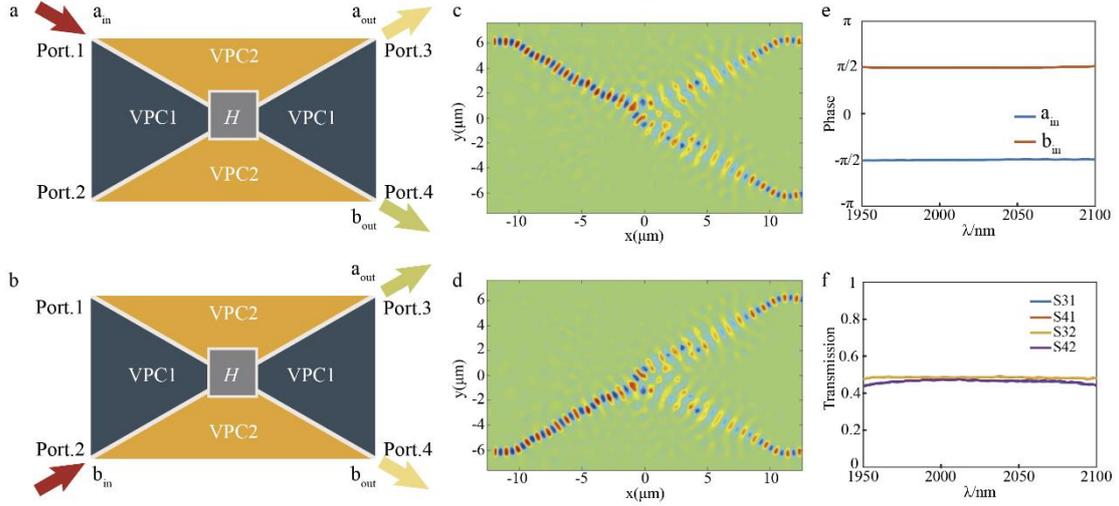

**Fig. S13. Hadamard Gate.** (**a** and **b**) Inverse-designed Hadamard gate under different input. (**c** and **d**) Simulation results of the Hadamard gate. The phase difference (**e**) and the transmission (**f**) of output port under different input conditions.

Furthermore, we have investigated the theoretical model and numerical simulations of an arbitrary single qubit gate R based on topological photonic crystals, as shown in Fig. S14a in the main text. For the path-encoded scheme, the R gate can be expressed as a series of matrices. Therefore, the evolution of the quantum state can be represented as:

$$|\varphi_{out}\rangle = \frac{1}{2}\begin{vmatrix}1 & 0 \\ 0 & e^{i\theta_3}\end{vmatrix}\begin{vmatrix}1 & 1 \\ 1 & -1\end{vmatrix}\begin{vmatrix}1 & 0 \\ 0 & e^{i\theta_2}\end{vmatrix}\begin{vmatrix}1 & 1 \\ 1 & -1\end{vmatrix}\begin{vmatrix}1 & 0 \\ 0 & e^{i\theta_1}\end{vmatrix}|\varphi_{in}\rangle \quad (S17)$$

when the quantum state is injected into the single qubit gate R, any unitary transformation can be applied by adjusting the phases $\theta_1$, $\theta_2$, $\theta_3$. We now present three examples to demonstrate its functionality through numerical simulation. In these three instances, the fixed phase in the R gate is set to $\theta=0$, $\pi/2$, or $\pi$, respectively, and the resulting quantum states are $|0\rangle$、$\frac{1}{\sqrt{2}}(|0\rangle+|1\rangle)$ and $|1\rangle$, where $|0\rangle$ and $|1\rangle$ are quantum states represented through different paths, as shown in Fig. S14b, c, and d. All three cases were simulated with a single photon excited from the upper waveguide, where the quantum state is defined as $|0\rangle$. Correspondingly, the quantum state in the lower waveguide is defined as $|1\rangle$. We found that the output quantum states indeed transform into the expected states after passing through the R gate. We map these output quantum states onto three points on the Bloch sphere. Any arbitrary output quantum state can also be generated by the R gate, corresponding to any point on the Bloch sphere. Therefore, the simulation results are in good agreement with the theoretical predictions. This indicates that the R gate can be realized in topological photonic crystals.

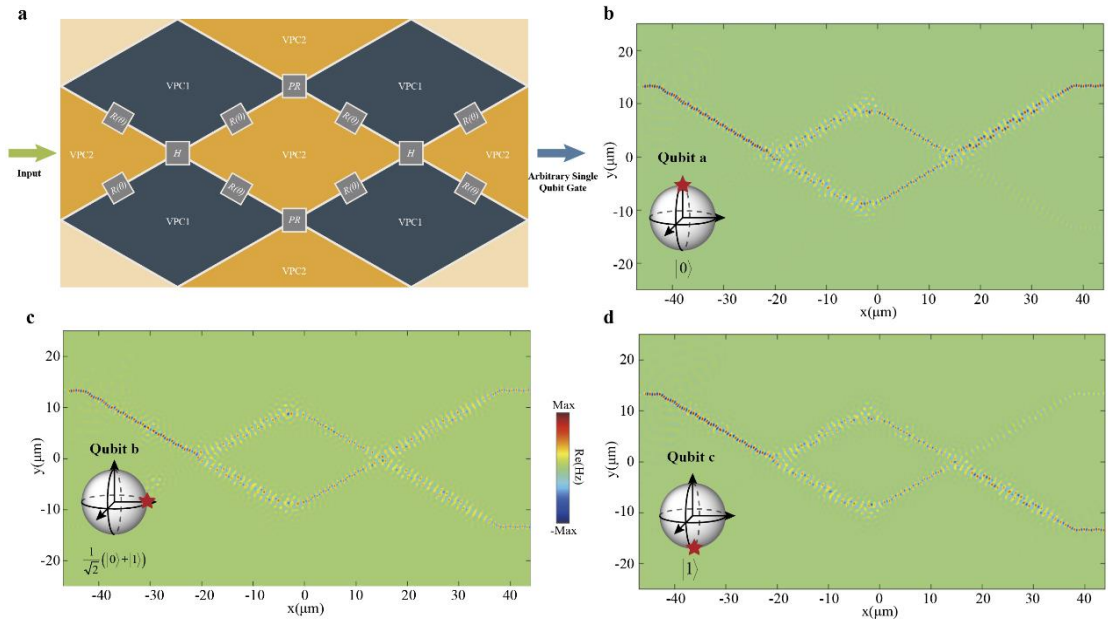

**Fig. S14. Arbitrary Single Qubit Gate.** (**a**) The arbitrary single-qubit gate consisted of phase Rz(θ) gates and Hadamard gates. When the phases of the phase gates are set as (**b**) 0, (**c**) π/2, (**d**) π, the simulation results of the arbitrary single qubit gate.

## Section 8. Topological Wavelength Division Multiplexing Device based on Inverse Design

Wavelength division multiplexing (WDM) was critically important for the application of topological photonic crystals. In topological photonic crystals, WDM could be achieved through inverse design structures[24] or precise design of photonic crystal unit structures[25]. However, these methods faced challenges such as high insertion loss, low extinction ratio, and a high dependency on fabrication precision.

Fig. S15 (a and b) were the functional diagram of the WDM device of the topological photonic crystal. Here we rewrote our objective function based on the previous router:

$$\begin{vmatrix} \mathbf{E}_{port\_3} \\ \mathbf{E}_{port\_4} \end{vmatrix} = \mathbf{T} \cdot \begin{vmatrix} \lambda_1 & \lambda_3 \\ \lambda_2 & \lambda_4 \end{vmatrix} \cdot \begin{vmatrix} \mathbf{E}_{port\_1} \\ \mathbf{E}_{port\_2} \end{vmatrix} \quad (S18)$$

The energy distribution of the device, at the corresponding wavelengths, was illustrated in Fig. S15 e. At a wavelength of 1950 nm, upon the input of energy from Port.1, a significant conversion to the kink states bounded in K' was observed, which was then channeled to Port.3 for output. Conversely, at 2000 nm, energy predominantly radiated from Port.4 in the kink states bounded in K. Our computational results indicated an impressive performance, as shown in Fig. S15 c. Within the first waveband (1940 ~ 1960 nm), the device exhibited an insertion loss (IL) of 0.41 dB (@1952 nm) and an extinction ratio (ER) of 24.01 dB (@1943 nm). As for the second waveband (1990 ~ 2010 nm), the device maintained a slight IL of 0.31 dB (@2005 nm) and an ER of 21.50 dB(@1998 nm).

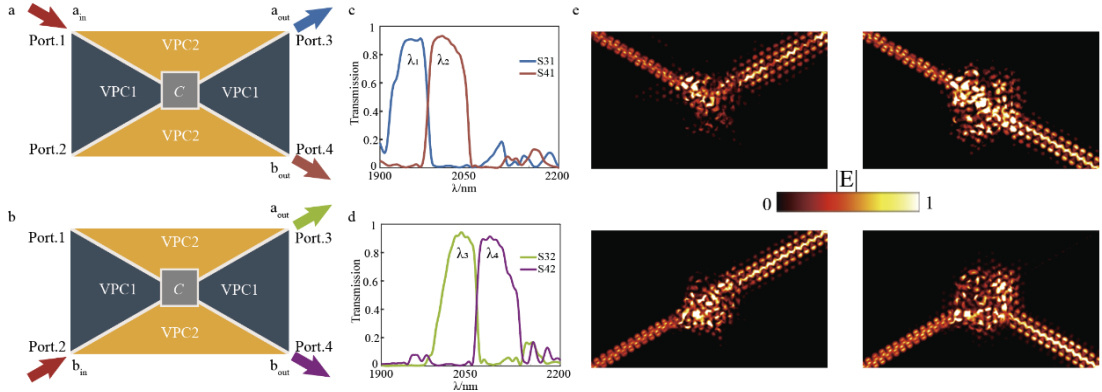

**Fig. S15. Topological Wavelength Division Multiplexing (WDM) device.** (**a** and **b**) Functional diagram of topological WDM device. (**c** and **d**) Simulated transmission spectra of the designed wavelength demultiplexer. (**e**) Simulated energy flow density plots of the device operating at different band

When the incident light has the opposite valley state, the device exhibits additional functionality. At a wavelength of 2050 nm, light entering through Port.2 is converted into the K' state and output from Port.3. Conversely, at a wavelength of 2100 nm, light incident on Port.2 was transformed into the K state and then output from Port.4, as illustrated in the Fig. S15 e. Device performance was validated through simulations. Within the third band, spanning 2030 to 2050 nm, the device exhibited an IL of 0.25 dB at 2035 nm and an ER of 24.58 dB at 2032 nm. Similarly, in the fourth band, which covers 2070 to 2090 nm, the device demonstrated an IL of 0.39 dB at 2082 nm and an ER of 20.87 dB at 2088 nm. To the best of our knowledge, these results are unparalleled in the current landscape of similar device types.

## Section 9. Spatial Light Coupler in Topological Photonic Crystal based on Inverse Design

Topological photonic crystals, characterized by their unique optical properties, hold significant promise for applications in free-space optical communications and Lidar[26]. Nevertheless, their utility is curtailed by challenges such as moderated directionality and poor diffraction-field tailoring capability, which hinder direct spatial coupling of patterns into these crystals. When a Gaussian beam is incident upon a topological photonic crystal, it predominantly results in energy loss through leaky modes, while simultaneously exciting two pseudo-spin modes propagating in divergent directions(Fig. S16a). The simulations demonstrate that the light field predominantly localizes at the incident area without forming effective propagation within the crystal, leading to an IL exceeding 45.5 dB, as shown in Fig. S16b. Through inverse design, we successfully converted spatial modes to pseudo-spin modes (Fig. S16c). The incident Gaussian beam on the device now predominantly localizes at the boundary of the topological waveguide, propagating unidirectionally, with the device exhibiting a loss of 7 dB (Fig. S16d). This performance is comparable to that of standard mid-infrared gratings. Furthermore, when light is incident from another angle, the signal is converted into another pseudospin mode and output from another direction, as shown in Fig. S16e. The simulation results show that the signal propagates in one direction and the device exhibits a loss of 10.5 dB. Taking advantage of the reciprocity of light, the designed device can also be used as a FSC for off-chip signal transmission.

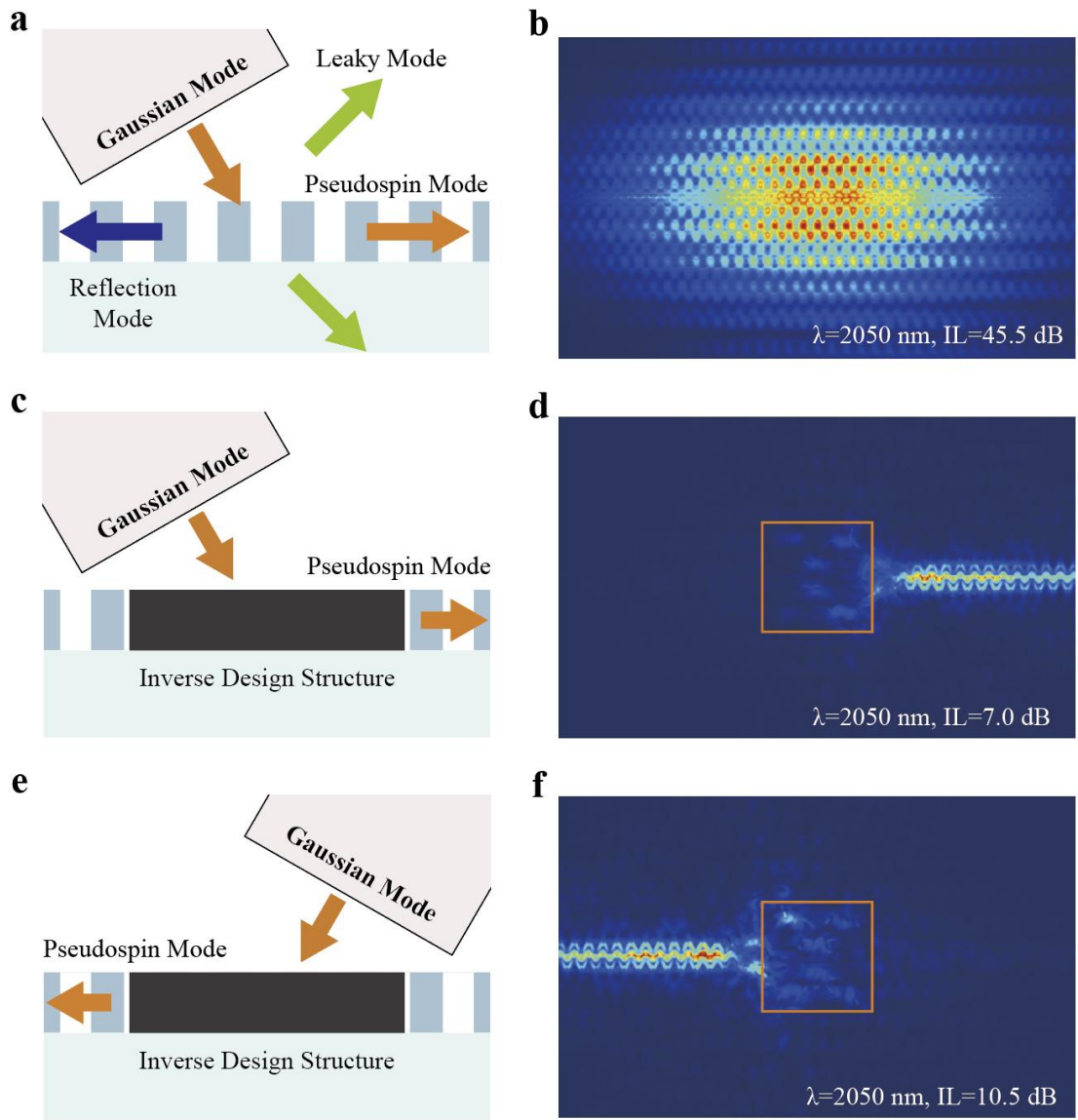

**Fig. S13. Spatial Light Coupler in Topological Photonic Crystal based on Inverse Design.** (**a**, **c** and **d**) Schematic diagram. (**b**, **d** and **f**) Light field distribution of the structure when a Gaussian light source is incident.